\documentclass{amsart}

\usepackage{amsmath}
\usepackage{amsfonts,amssymb}
\usepackage{mathrsfs}

\newtheorem{theorem}{Theorem}[section]
\newtheorem{proposition}{Proposition}[section]
\newtheorem{lemma}[theorem]{Lemma}
\newtheorem{corollary}[theorem]{Corollary}
\theoremstyle{definition}
\newtheorem{definition}[theorem]{Definition}
\newtheorem{example}[theorem]{Example}
\newtheorem{remark}[theorem]{Remark}
\numberwithin{equation}{section}

\begin{document}

\title{On the separation principle of quantum control}

\author{Luc Bouten}
\address{Luc Bouten \\
Physical Measurement and Control 266-33 \\
California Institute of Technology \\
Pasadena, CA 91125 \\
USA}
\email{bouten@its.caltech.edu}

\author{Ramon van Handel}
\address{Ramon van Handel \\
Physical Measurement and Control 266-33 \\
California Institute of Technology \\
Pasadena, CA 91125 \\
USA}
\email{ramon@its.caltech.edu}
\thanks{This work was supported in part by the NSF under contract number
PHY-0456720, and in part by the ARO under contract number DAAD19-03-1-0073.}

\begin{abstract} 
It is well known that quantum continuous observations and nonlinear 
filtering can be developed within the framework of the quantum stochastic
calculus of Hudson-Parthasarathy.  The addition of real-time feedback
control has been discussed by many authors, but the foundations of the 
theory still appear to be relatively undeveloped.  Here we introduce 
the notion of a controlled quantum flow, where feedback is taken into 
account by allowing the coefficients of the quantum stochastic 
differential equation to be adapted processes in the observation algebra.  
We then prove a separation theorem for quantum control: the admissible 
control that minimizes a given cost function is a memoryless function of 
the filter, provided that the associated Bellman equation has a 
sufficiently regular solution.  Along the way we obtain results on 
existence and uniqueness of the solutions of controlled quantum filtering 
equations and on the innovations problem in the quantum setting.
\end{abstract}

\maketitle

%%%%%%%%%%%%%%%%%%%%%%%%%%%%%%%%%%%%%%%%%%%

\section{Introduction}

Quantum feedback control is a branch of stochastic control theory that
takes into account the inherent uncertainty in quantum systems. Though
quantum stochastic control was first investigated in the 1980s in the
pioneering papers of Belavkin \cite{Bel83,Bel88} it is only recently that
this has become a feasible technology, as demonstrated by recent
laboratory experiments in quantum optics \cite{AASDM02,GSM04}.  On the
other hand, modern computing and sensing technology are rapidly reaching 
a level of miniaturization and sensitivity at which inherent quantum
uncertainties can no longer be neglected.  The development of control
theoretic machinery for the design of devices that are robust in presence
of quantum uncertainty could thus have important implications for a
future generation of precision technology.

Though not as mature as their classical counterparts, mathematical tools
for quantum stochastic analysis have been extensively developed over the
last two decades following the introduction of quantum stochastic calculus
by Hudson and Parthasarathy \cite{HuP84}.  Quantum stochastic differential
equations are known to provide accurate Markov models of realistic quantum
systems, particularly the atomic-optical systems used in quantum optics,
and continuous-time optical measurements are also accurately described
within this framework.  Furthermore, nonlinear filtering theory for
quantum systems has been extensively developed \cite{Bel92b,BGM04,BvH05}
and provides a suitable notion of conditioning for quantum systems.  
Nonetheless the theory of quantum control is still very much in its
infancy, and despite the large body of literature on classical stochastic
control only a few rigorous results are available in the quantum case.  
Our goal here is to make a first step in this direction by proving a
quantum version of a simple but important theorem in classical control 
theory: a separation theorem for optimal stochastic controls.

To set the stage for the remainder of the article, let us demonstrate the 
idea with a simple (but important) example.  We consider an atom in 
interaction with the vacuum electromagnetic field; the atom can emit 
photons into the field, and we allow ourselves to control the strength of 
a fixed atomic Hamiltonian.  The system dynamics is given by the quantum 
stochastic differential equation (QSDE)
$$
	dj_t(X)= u(t)\,j_t(i[H,X])\,dt + j_t(\mathcal{L}[X])\,dt +
	j_t([X,L])\,dA_t^*+j_t([L^*,X])\,dA_t
$$
where $j_t(X)$ denotes the atomic observable $X$ at time $t$, and for now 
$u(t)$ is a deterministic control function (i.e.\ an open loop control).  
If we perform homodyne detection in the field, we observe the stochastic 
process $Y_t$ given by
$$
	dY_t = j_t(L+L^*)\,dt+dA_t^*+dA_t.
$$
We now have a system-observation pair as in classical stochastic control.  
Inspired by results in classical control theory, we begin by finding a 
recursive equation for $\pi_t(X)=\mathbb{P}(j_t(X)|\mathcal{Y}_t)$, the 
conditional expectation of the atomic observable $X$ at time $t$, given 
the observations $Y_s$ up to time $t$.  One obtains the nonlinear filter
\begin{multline*}
	d\pi_t(X)=u(t)\,\pi_t(i[H,X])\,dt+\pi_t(\mathcal{L}[X])\,dt \\
	+(\pi_t(XL+L^*X)-\pi_t(L+L^*)\pi_t(X))\,(dY_t-\pi_t(L+L^*)\,dt).
\end{multline*}
A crucial property of the conditional expectation is that the expectation
of $\pi_t(X)$ equals the expectation of $j_t(X)$, i.e.\ 
$\mathbb{P}(\pi_t(X))=\mathbb{P}(j_t(X))$.  Suppose we pose as our control 
goal the preparation of an atomic state with particular properties, e.g.\ 
we wish to find a control $u(t)$ such that after a long time
$\mathbb{P}(j_t(X))=\mathbb{P}_f(X)$ for some target state $\mathbb{P}_f$.
Then it is sufficient to design a control that obeys this property for the 
filter $\pi_t(X)$.  The advantage of using the filter is that $\pi_t(X)$ 
is only a function of the observations, and hence is always accessible to 
us, unlike $j_t(X)$ which is not directly observable.  This approach was 
taken e.g.\ in \cite{HSM05,MvH05}.

We immediately run into technical problems, however, as we have assumed in
the derivation of the filtering equation that $u(t)$ is a deterministic
function whereas state preparation generally requires the use of a
feedback control.  One can of course simply replace the deterministic
function $u(t)$ in the filter with some (feedback) function of the
observation history, which is the approach generally taken in the
literature, but does this new equation actually correspond to the
associated controlled quantum system?  The first issue that we resolve in
this paper is to show that if $u(t)$ is replaced by some function of the
observation history in the equation for $j_t(X)$, then the associated
filtering equation corresponds precisely (as expected)  to the open loop
filtering equation obtained without feedback where the control $u(t)$ is
replaced by the same function of the observations.

The separation of the feedback control strategy into a filtering step and
a control step, as suggested above, is a desirable situation, as the
filter can be calculated recursively and hence the control strategy is not
difficult to implement.  It is not obvious, however, that such a
separation is always possible.  Rather than taking state preparation as
the control objective, consider the optimal control problem in which the
control goal is to find a control strategy that minimizes a suitably
chosen cost function.  This is a common choice in control theory, and in
general the cost function can even be expressed in terms of the nonlinear
filter.  However, it is not at all obvious that the {\it optimal} control
at time $t$ only depends on $\pi_t(X)$; in principle, the control could
depend on the entire past history of the filter or even on some aspect of
the observation process that is not captured by the filtering equation!  
The implementation of such a control would be awkward, as it would require
the controller to have sufficient memory to store the entire observations
history and enough resources to calculate an appropriate functional
thereof.

Optimal control problems are often approached through the method of
dynamic programming, which provides a candidate control strategy in
separated form.  We will show that if we can find a separated control
strategy that satisfies the dynamic programming equations, then this
strategy is indeed optimal even with respect to all non-separated
strategies.  Thus the fortunate conclusion is that even in the case of
optimal controls we generally do not need to worry about non-separated
control strategies.  This establishes a foundation for the {\it
separation principle} of quantum control, by which we mean that as a rule
of thumb the design of quantum feedback controls can be reduced to a
separate filtering step and a control step.

On the technical side, our treatment of quantum filtering proceeds by
means of the reference probability approach \cite{BvH05} which is inspired
by the approach of Zakai \cite{Zak69} in classical nonlinear filtering.  
Our treatment of the separation theorem is directly inspired by the 
classic papers of Wonham \cite{Won68} and Segall \cite{Seg77}.
A fully technical account of the results in this article will be presented 
in \cite{BvH05b}, and we apologize to the reader for the liberally 
sprinkled references to that paper.  Here we will mostly neglect domain 
issues and similar technicalities, while we focus our attention on 
demonstrating the results announced above.

This paper is organized as follows.  In section \ref{sec:QP} we briefly
recall some of the basic ideas of quantum probability theory, and we
develop the reference probability approach to quantum filtering without
feedback.  In section \ref{sec:controlqf} we introduce the notion of a
controlled quantum flow and show that for such models the controlled
filter takes the expected form.  In section \ref{sec:innovations} we
convert the filtering equations into classical stochastic differential
equations and study their sample path properties; as a corollary, we
obtain some results on the innovations problem.  Finally, in section
\ref{sec separation} we introduce the optimal control problem and prove
a separation theorem.

\section{Quantum probability and filtering}
\label{sec:QP}

The purpose of this section is to briefly remind the reader of the basic 
ideas underlying quantum probability and filtering.  For a more thorough 
introduction we refer to \cite{BvH05} and the references therein.

A quantum probability space $(\mathcal{A},\mathbb{P})$ consists of a von
Neumann algebra $\mathcal{A}$, defined on some underlying Hilbert space
$\mathscr{H}$, and a normal state $\mathbb{P}$.  If $\mathcal{A}$ is
commutative then this definition is essentially identical to the usual
definition in classical probability theory: indeed, the spectral theorem
then guarantees that for some measure space $(\Omega,\Sigma,\mu)$ there
exists a $^*$-isomorphism $\iota:\mathcal{A}\to
L^\infty(\Omega,\Sigma,\mu)$ such that $\mathbb{P}(A)=\mathbb{E}_{\bf
P}(\iota(A))$ for all $A\in\mathcal{A}$, where $\mathbb{E}_{\bf P}$
denotes the expectation w.r.t.\ the probability measure ${\bf
P}\ll\mu$.\footnote{
	We will also denote the $^*$-isomorphism as 
	$\iota:(\mathcal{A},\mathbb{P})\to 
	L^\infty(\Omega,\Sigma,\mu,{\bf P})$; this means that the
	null sets are quotiented w.r.t.\ the measure $\mu$,
	whereas $\mathbf{P}\ll\mu$ is the image of the state $\mathbb{P}$.
}  Thus any self-adjoint element of $\mathcal{A}$ represents a
bounded random variable (observable).  In quantum models $\mathcal{A}$ is
noncommutative, but in each realization we are only allowed to measure a
commuting set of observables which generate a commutative von Neumann
algebra $\mathcal{C}$.  Hence if we fix a set of (commuting) observations
to be performed in every realization of the experiment, then many
computations can be reduced to classical probability theory.

To define the notion of a conditional expectation in quantum probability, 
we now simply ``pull back'' the associated notion from classical 
probability theory:

\begin{definition}[Conditional expectation]\label{def:condex}
Let $(\mathcal{A},\mathbb{P})$ be a quantum probability space and
$\mathcal{C}\subset\mathcal{A}$ be a commutative von Neumann algebra.
Define the commutant $\mathcal{C}'=\{A\in\mathcal{A}:AC=CA~\forall
C\in\mathcal{C}\}$. The map
$\mathbb{P}(\cdot|\mathcal{C}):\mathcal{C}'\to\mathcal{C}$ is called (a
version of) the conditional expectation onto $\mathcal{C}$ if
$\mathbb{P}(\mathbb{P}(A|\mathcal{C})C)=\mathbb{P}(AC)$ for all
$A\in\mathcal{C}'$, $C\in\mathcal{C}$. 
\end{definition}

Let us clarify the statement that this is nothing more than a classical 
conditional expectation.  Let $A\in\mathcal{C}'$ be self-adjoint; then 
$\mathcal{C}_A={\rm vN}(A,\mathcal{C})$, the von Neumann algebra generated 
by $A$ and $\mathcal{C}$, is a commutative algebra, and the spectral 
theorem gives a $^*$-isomorphism $\iota_A$ to some 
$L^\infty(\Omega_A,\Sigma_A,\mu_A,{\bf P}_A)$.  But then we can simply 
calculate the classical conditional expectation and pull it back to the 
algebra: $\mathbb{P}(A|\mathcal{C})=\iota_A^{-1}(\mathbb{E}_{{\bf 
P}_A}(\iota_A(A)|\sigma(\iota_A(\mathcal{C}))))$.  This is in fact 
identical to Definition \ref{def:condex}, and can be extended to any
$A$ by writing is as $B+iC$ where $B,C$ are self-adjoint.

The definition we have given is less general than the usual definition
\cite{Tak71}.  In particular, we only allow conditioning onto a
commutative algebra $\mathcal{C}$ from its commutant $\mathcal{C}'$.  For
statistical inference purposes (filtering) this is in fact sufficient: we
wish to condition on the set of measurements made in a single realization
of an experiment, hence they must commute; and it only makes sense to
condition observables that are compatible with the observations already
made, otherwise the conditional statistics would not be detectable by any
experiment.  Definition \ref{def:condex} has the additional advantage that
existence, uniqueness, and all the basic properties can be proved by
elementary means \cite{BvH05}.

We list some of the most important properties of the conditional
expectation: it exists, is a.s.\ unique (the difference between any two
versions is zero with unit probability), and satisfies the least-squares
property $|||A-\mathbb{P}(A|\mathcal{C})|||\le |||A-C|||$,
$|||X|||^2=\mathbb{P}(X^*X)$ for all $C\in\mathcal{C}$. Moreover we have
(up to a.s.\ equivalence) linearity, positivity, invariance of the state
$\mathbb{P}(\mathbb{P}(A|\mathcal{C}))=\mathbb{P}(A)$, the module property
$\mathbb{P}(AB|\mathcal{C})=B\,\mathbb{P}(A|\mathcal{C})$ for
$B\in\mathcal{C}$, the tower property
$\mathbb{P}(\mathbb{P}(A|\mathcal{B})|\mathcal{C})=
\mathbb{P}(A|\mathcal{C})$ if $\mathcal{C}\subset\mathcal{B}$, etc.  
Finally, we have the following Bayes-type formula:

\begin{lemma}[Bayes formula \cite{BvH05}]\label{lem:bayes}
Let $(\mathcal{A},\mathbb{P})$ be a quantum probability space and
$\mathcal{C}\subset\mathcal{A}$ be a commutative von Neumann algebra.
Furthermore, let $V\in\mathcal{C}'$ be such that $V^*V>0$,
$\mathbb{P}(V^*V)=1$.  
Then we can define a new state on $\mathcal{C}'$ by $\mathbb{Q}(A) = 
\mathbb{P}(V^*AV)$ and we have $\mathbb{Q}(X|\mathcal{C}) = 
\mathbb{P}(V^*XV|\mathcal{C})\,/\,\mathbb{P}(V^*V|\mathcal{C})$ for $X\in 
\mathcal{C}'$.
\end{lemma}

Up to this point we have only dealt with bounded operators.  In the
framework of quantum stochastic calculus, however, we unavoidably have to
deal with unbounded operators that are affiliated to, not elements of, the
various algebras mentioned above.  As announced in the introduction, we
largely forgo this issue here and we claim that all the results above (and
below) can be extended to a sufficiently large class of unbounded
operators.  We refer to \cite{BvH05b} for a complete treatment.

Let us now introduce a class of quantum models that we will consider in
this paper.  The model consists of an initial system, defined on a
finite-dimensional Hilbert space $\mathscr{H}_0$, in interaction with an
external (e.g.\ electromagnetic) field that lives on the usual Boson Fock
space $\Gamma=\Gamma_s(L^2([0,T]))$.  We will always work on a finite time
horizon $[0,T]$ and we have restricted ourselves for simplicity to a
single channel in the field (the case of multiple channels presents no
significant complications; it can be treated in the same manner on a case
by case basis \cite{BvH05}.  A completely general theory can also be set 
up, e.g.\ \cite{Bel92b}, but the required notations seem unnecessarily 
complicated.)  
Throughout we place ourselves on the quantum probability space
$(\mathcal{A},\mathbb{P})$ where $\mathcal{A}=\mathcal{B}\otimes\mathcal{W}$,
$\mathcal{B}=B(\mathscr{H}_0)$, $\mathcal{W}=B(\Gamma)$, and
$\mathbb{P}=\rho\otimes\phi$ for some state $\rho$ on $\mathcal{B}$ and
the vacuum state $\phi$ on $\mathcal{W}$.  We will use the standard
notation $\Gamma_{t]}=\Gamma_s(L^2([0,t]))$,
$\mathcal{W}_{t]}=B(\Gamma_{t]})$, etc.  For $f\in L^2([0,T])$ we denote 
by $e(f)\in\Gamma$ the corresponding exponential vector, by $\Phi=e(0)$ 
the vacuum vector, and by $A_t$, $A_t^*$ and $\Lambda_t$ the fundamental
noises.  The reader is referred to \cite{HuP84,Bia95,Mey93,Par92} for
background on quantum stochastic calculus.

For the time being we will not consider feedback control---we extend to 
this case in section \ref{sec:controlqf}.  Without feedback, the 
interaction between the initial system and the field is given by the 
unitary solution of the Hudson-Parthasarathy QSDE
$$
	U_t=I
	+\int_0^tL_sU_s\,dA_s^*
	-\int_0^tL_s^*S_sU_s\,dA_s
	+\int_0^t(S_s-I)U_s\,d\Lambda_s
	-\int_0^t(
		iH_s+\tfrac{1}{2}L_s^*L_s
	)U_s\,ds.
$$
Here $L_t$, $S_t$ and $H_t$ are bounded processes of operators in
$\mathcal{B}$, $S_t$ is unitary and $H_t$ is self-adjoint.  Without
external controls the processes will usually be chosen to be
time-independent, or we can imagine that e.g.\ the Hamiltonian $H_t=u(t)H$
is modulated by some {\it deterministic (open loop)} scalar control
$u(t)$.  In addition we specify an output noise that will be measured in
the field; it takes the general form
$$
	Z_t=\int_0^t \lambda_s\,d\Lambda_s+
		\int_0^t\alpha_s\,dA_s^*+
		\int_0^t\alpha_s^*\,dA_s
$$
where $\lambda:[0,T]\to\mathbb{R}$ and $\alpha:[0,T]\to\mathbb{C}$ are
bounded scalar functions.  Together $U_t$ and $Z_t$ provide a full
description of a filtering problem: any initial system observable
$X\in\mathcal{B}$ is given at time $t$ by the flow $j_t(X)=U_t^*XU_t$,
whereas the observation process that appears on our detector is given by
$Y_t=U_t^*Z_tU_t$.  Using the quantum It\^o rules, we obtain the explicit 
expressions
\begin{multline}\label{eq:syssys}
	dj_t(X)=
	j_t(i[H_t,X]
		+L_t^*XL_t-\tfrac{1}{2}\{L_t^*L_tX+
		XL_t^*L_t\})\,dt \\
	+j_t(S_t^*[X,L_t])\,dA_t^*
	+j_t([L_t^*,X]S_t)\,dA_t
	+j_t(S_t^*XS_t-X)\,d\Lambda_t,
\end{multline}
\begin{multline}\label{eq:sysobs}
	dY_t = \lambda_t\,d\Lambda_t
	+j_t\left(
		S_t^*(\alpha_t+\lambda_tL_t)
	\right)\,dA_t^*
	+j_t\left(
		(\alpha_t^*+\lambda_tL_t^*)S_t
	\right)\,dA_t \\
	+j_t\left(
		\lambda_tL_t^*L_t+\alpha_t^*L_t+\alpha_tL_t^*
	\right)\,dt.
\end{multline}
This ``system-theoretic'' description in terms of the system-observations 
pair (\ref{eq:syssys}) and (\ref{eq:sysobs}) is closest in spirit to the 
usual description of filtering and control problems in classical control 
theory.  We will not explicitly use this representation, however.

We now turn to the filtering problem, i.e.\ the problem of finding an
explicit representation for the conditional state $\pi_t(X)=
\mathbb{P}(j_t(X)|\mathcal{Y}_t)$, $X\in\mathcal{B}$, where
$\mathcal{Y}_t={\rm vN}(Y_s:0\le s\le t)$ is the von Neumann algebra
generated by the observations up to time $t$.  Before we can go down this
road we must prove that $\pi_t(X)$ is in fact well defined according to
Definition \ref{def:condex}.  This is guaranteed by the following
proposition.

\begin{proposition}[Nondemolition property]\label{pro:nondemolition}
	The observation process $Y_t$ satisfies the self-nondemolition 
	condition, i.e.\ $\mathcal{Y}_t$ is commutative for all $t\in[0,T]$, 
	and is nondemolition with respect to the flow, i.e.\ 
	$j_t(X)\in\mathcal{Y}_t'$ for all $X\in\mathcal{B}$ and $t\in[0,T]$.
\end{proposition}

\begin{proof}
Let $\mathcal{Z}_t={\rm vN}(Z_s:0\le s\le t)$.  We begin by showing that 
$\mathcal{Z}_t$ is a commutative algebra for all $t\in[0,T]$.  To this 
end, define
$$
	Z(c,d)=\int_0^T c_s\,d\Lambda_s+
		\int_0^T d_s\,dA_s^*+
		\int_0^T d_s^*\,dA_s
$$
so that $Z_t=Z(\lambda\chi_{[0,t]},\alpha\chi_{[0,t]})$.  Using the 
quantum It\^o rules, we obtain
\begin{multline*}
	[Z(C,D),Z(c,d)]=\\
	\int_0^T(C_sd_s-c_sD_s)\,dA_s^*
	+\int_0^T(D^*_sc_s-d^*_sC_s)\,dA_s
	+\int_0^T(D^*_sd_s-d^*_sD_s)\,dt.
\end{multline*}
But then we obtain $[Z_t,Z_{t'}]=0$ for all $t,t'\in[0,T]$ by setting
$$
	C_s=\lambda_s\chi_{[0,t]}(s),\quad
	c_s=\lambda_s\chi_{[0,t']}(s),\quad
	D_s=\alpha_s\chi_{[0,t]}(s),\quad
	d_s=\alpha_s\chi_{[0,t']}(s).
$$
We conclude that the process $Z_t$ generates a commutative algebra.

Next, we claim that $U_s^*Z_sU_s=U_t^*Z_sU_t$ for all $s\le t\in[0,T]$.  
To see this, let $E\in\mathcal{Z}_{s}$ be an arbitrary projection operator 
in the range of the spectral measure of $Z_s$.  Using the quantum It\^o 
formula, we obtain
\begin{multline*}
	j_t(E)=j_{s}(E)+
	\int_s^t j_\sigma(i[H_\sigma,E]
		+L_\sigma^*EL_\sigma-\tfrac{1}{2}\{L_\sigma^*L_\sigma E+
		EL_\sigma^*L_\sigma\})\,d\sigma \\
	+\int_s^t j_\sigma(S_\sigma^*[E,L_\sigma])\,dA_\sigma^*
	+\int_s^t j_\sigma([L_\sigma^*,E]S_\sigma)\,dA_\sigma
	+\int_s^t j_\sigma(S_\sigma^*ES_\sigma-E)\,d\Lambda_\sigma
\end{multline*}
where $j_t(X)=U_t^*XU_t$.  But by construction $E$ commutes with all 
$H_\sigma,L_\sigma,S_\sigma$, hence we obtain $j_t(E)=j_s(E)$.  As this 
holds for all spectral projections $E$ of $Z_s$, the assertion follows.
We conclude that $Y_t=U_T^*Z_tU_T$ for all $t\in[0,T]$.  But then 
$\mathcal{Y}_t=U_T^*\mathcal{Z}_tU_T$, and as $\mathcal{Z}_t$ is
commutative $\mathcal{Y}_t$ must be as well.

It remains to prove the nondemolition condition.  To this end, note that
$j_t(X)\in U_t^*\mathcal{B}U_t$ and $\mathcal{Y}_t=U_t^*\mathcal{Z}_tU_t$.  
But $\mathcal{Z}_t$ consists of elements in $\mathcal{W}$, which clearly
commute with every element in $\mathcal{B}$.  The result follows 
immediately.
\end{proof}

To obtain an explicit representation for the filtering equation we are
inspired by the classical reference probability method of Zakai
\cite{Zak69}.  The idea of Zakai is to introduce a change of measure such
that under the new (reference) measure the observation process is a
martingale which is independent of the system observables, which 
significantly simplifies the calculation of conditional expectations.  The 
Bayes formula then relates the conditional expectations under the 
reference measure and the actual measure.  We will follow a similar route 
and make use of the quantum Bayes formula of Lemma \ref{lem:bayes}.  The 
main difficulty is the choice of an appropriate change of measure operator $V$.

In the classical reference probability method the change of measure is
obtained from Girsanov's theorem.  Unfortunately, there is no satisfactory
noncommutative analog of the Girsanov theorem; even though Girsanov-like
expressions can be obtained, they do not give rise to a change of state
that lies in the commutant of the observations as required by Lemma
\ref{lem:bayes}.  A different naive choice would be something like
$\mathbb{R}(X)=\mathbb{P}(U_TXU_T^*)$, so that $Y_t$ under $\mathbb{R}$
has the same statistics as the martingale $Z_t$ under $\mathbb{P}$, but
once again $U_T$ does not commute with the observations.  However, the
latter idea (in a slightly modified form) can be ``fixed'' to work:  
starting from a change of state that is the solution of a QSDE, we can
modify the QSDE somewhat so that the resulting solution still defines the
same state but has the desired properties.  This trick appears to have 
originated in a paper by Holevo \cite{Hol90}.  We state it here in the 
following form.

\begin{lemma}\label{lem:holevo}
        Let $C_t,D_t,F_t,G_t,\tilde C_t,\tilde F_t$ be bounded processes, 
	and let
	\begin{equation*}
	\begin{split}
                &dV_t=\{C_t\,d\Lambda_t+D_t\,dA_t^*+F_t\,dA_t+G_t\,dt
			\}V_t, \\
                &d\tilde V_t=\{\tilde C_t\,d\Lambda_t+
			D_t\,dA_t^*+\tilde F_t\,dA_t+G_t\,dt
			\}\tilde V_t,\quad V_0=\tilde V_0.
	\end{split}
	\end{equation*}
	Then $\mathbb{P}(V_t^*XV_t)=\mathbb{P}(\tilde V_t^*X\tilde V_t)$
        for all $X\in\mathcal{B}\otimes\mathcal{W}$.
\end{lemma}

\begin{proof}
As any state $\rho$ on $\mathcal{B}$ is a convex combination of vector
states, it is sufficient to prove the Lemma for any vector state
$\rho(B)=\langle v,Bv\rangle$, $v\in\mathscr{H}_0$.  Hence it suffices
to prove that $\langle V_t\,v\otimes\Phi,XV_t\,v\otimes\Phi\rangle=
\langle\tilde V_t\,v\otimes\Phi,X\tilde V_t\,v\otimes\Phi\rangle$ for any
$X\in\mathcal{B}\otimes\mathcal{W}$ and $v\in\mathscr{H}_0$, as
$\mathbb{P}=\rho\otimes\phi$.  But clearly this would be implied by
$V_t\,v\otimes\Phi=\tilde V_t\,v\otimes\Phi$ $\forall v\in\mathscr{H}_0$.
Let us prove that this is in fact the case.

As all the coefficients of the QSDE for $V_t,\tilde V_t$ are bounded 
processes both $V_t$ and $\tilde V_t$ have unique solutions.  Consider the 
quantity
$$
        \|(V_t-\tilde V_t)\,v\otimes\Phi\|^2=
        \langle (V_t-\tilde V_t)\,v\otimes\Phi,
        (V_t-\tilde V_t)\,v\otimes\Phi\rangle.
$$
Using the quantum It\^o rule we obtain (see e.g.\ \cite{Mey93})
\begin{multline*}
        \|(V_t-\tilde V_t)\,v\otimes\Phi\|^2=
        \int_0^t
                \langle (V_s-\tilde V_s)\,v\otimes\Phi,
                (G_s+G_s^*)(V_s-\tilde V_s)\,v\otimes\Phi\rangle
        \,ds
\\
        +\int_0^t
                \langle D_s(V_s-\tilde V_s)\,v\otimes\Phi,
                D_s(V_s-\tilde V_s)\,v\otimes\Phi\rangle
        \,ds.
\end{multline*}
Note that the last integrand can be expressed as
$$
        \|D_s(V_s-\tilde V_s)\,v\otimes\Phi\|^2\le
        \left[\sup_{t\in[0,T]}\|D_t\|^2\right]
        \|(V_s-\tilde V_s)\,v\otimes\Phi\|^2.
$$
To deal with the first integrand, note that $G_s+G_s^*$ are self-adjoint 
bounded operators.  Denote by $G^+_s$ the positive part
of $G_s+G_s^*$, and by $K^+_s$ the square root of $G^+_s$
(i.e.\ $G^+_s=K^{+*}_sK^+_s$).  Then
$$
        \langle (V_s-\tilde V_s)\,v\otimes\Phi,
        (G_s+G_s^*)(V_s-\tilde V_s)\,v\otimes\Phi\rangle\le
        \|K^+_s(V_s-\tilde V_s)\,v\otimes\Phi\|^2.
$$
But as $G_s+G_s^*$ is a bounded process, so is $K^+_s$ and we have
$$
        \|K^+_s(V_s-\tilde V_s)\,v\otimes\Phi\|^2\le
        \left[
                \sup_{t\in[0,T]}\|K^+_t\|^2
        \right] \|(V_s-\tilde V_s)\,v\otimes\Phi\|^2.
$$
Thus we obtain
$$
        \|(V_t-\tilde V_t)\,v\otimes\Phi\|^2\le
        C\int_0^t \|(V_s-\tilde V_s)\,v\otimes\Phi\|^2\,ds
$$
were by boundedness of $D_t$ and $K^+_t$
$$
        C=\sup_{t\in[0,T]}\|K^+_t\|^2+
        \sup_{t\in[0,T]}\|D_t\|^2<\infty.
$$
But then by Gronwall's lemma $\|(V_t-\tilde V_t)\,v\otimes\Phi\|=0$,
and the Lemma is proved.
\end{proof}

We are now in the position to prove the main filtering theorem: a quantum 
version of the Kallianpur-Striebel formula.  We consider separately two 
versions of the theorem for diffusive and counting observations, 
respectively (but note that the former also covers some cases with 
counting observations if $\lambda_t\ne 0$.)

\begin{theorem}[Kallianpur-Striebel formula, diffusive case]\label{thm:KSdiff}
Suppose that $|\alpha_t|>0$ for all $t\in[0,T]$.  Let $V_t$ be the 
solution of the QSDE
$$
	V_t=I+\int_0^t
		\alpha_s^{-1}L_sV_s
	\,(\lambda_s\,d\Lambda_s+\alpha_s\,dA_s^*+\alpha_s^*\,dA_s)-
	\int_0^t(iH_s+\tfrac{1}{2}L_s^*L_s)V_s\,ds.
$$
Then $V_t$ is affiliated to $\mathcal{Z}_t'$ for every $t\in[0,T]$, and
we have the representation $\pi_t(X)=\sigma_t(X)/\sigma_t(I)$ with
$\sigma_t(X)=U_t^*\mathbb{P}(V_t^*XV_t|\mathcal{Z}_t)U_t$,
$X\in\mathcal{B}$.
\end{theorem}

\begin{theorem}[Kallianpur-Striebel formula, counting case]\label{thm:KScount}
Suppose that $|\lambda_t|>0$ for all $t\in[0,T]$.  Let $E_t$ be the 
solution of the QSDE
$$
	E_t=I+
	\int_0^t\lambda_s^{-1}(1-\alpha_s)E_s\,dA_s^*-
	\int_0^t\lambda_s^{-1}(1-\alpha_s^*)E_s\,dA_s-
	\frac{1}{2}\int_0^t
		\lambda_s^{-2}|1-\alpha_s|^2E_s\,ds
$$
and let $C_t=E_t^*Z_tE_t$, $\mathcal{C}_t={\rm vN}(C_s:0\le s\le t)
=E_t^*\mathcal{Z}_tE_t$, so that
$$
	C_t=A_t^*+A_t+\int_0^t\lambda_s\,d\Lambda_s+
	\int_0^t\lambda_s^{-1}(1-|\alpha_s|^2)\,ds.
$$
Define $V_t$ as the solution of the QSDE
\begin{multline*}
	V_t=I+\int_0^t
		(L_s-\lambda_s^{-1}(1-\alpha_s))V_s
	\,(\lambda_s\,d\Lambda_s+dA_s^*+dA_s) \\
	-\int_0^t(iH_s+\tfrac{1}{2}L_s^*L_s
		+\tfrac{1}{2}\lambda_s^{-2}|1-\alpha_s|^2
		-\lambda_s^{-1}(1-\alpha_s^*)L_s)V_s\,ds.
\end{multline*}
Then $V_t$ is affiliated to $\mathcal{C}_t'$ for every $t\in[0,T]$, and we 
have the representation $\pi_t(X)=\sigma_t(X)/\sigma_t(I)$ with
$\sigma_t(X)=U_t^*E_t\mathbb{P}(V_t^*XV_t|\mathcal{C}_t)E_t^*U_t$,
$X\in\mathcal{B}$.
\end{theorem}

\begin{proof}[Proof of Theorem \ref{thm:KSdiff}]
We begin by using a general transformation property of the quantum 
conditional expectation.  Let $U$ be a unitary operator and define a new 
state $\mathbb{Q}(X)=\mathbb{P}(U^*XU)$.  Then 
$\mathbb{P}(U^*XU|U^*\mathcal{C}U)=U^*\mathbb{Q}(X|\mathcal{C})U$ 
provided that $\mathcal{C}$ is commutative and $X\in\mathcal{C}'$.  The 
statement is easily verified by direct application of Definition 
\ref{def:condex}.  The reason we wish to perform such a transformation is 
that in order to apply Lemma \ref{lem:holevo}, it will be more convenient 
if we condition with respect to $\mathcal{Z}_t$ rather than 
$\mathcal{Y}_t$.  From this point on, we fix a $t\in[0,T]$ and define the 
new state $\mathbb{Q}(X)=\mathbb{P}(U_t^*XU_t)$.  Evidently 
$\pi_t(X)=U_t^*\mathbb{Q}(X|\mathcal{Z}_t)U_t$.

We would now like to apply Lemma \ref{lem:bayes} to
$\mathbb{Q}(X|\mathcal{Z}_t)$.  Note that by Lemma \ref{lem:holevo} we 
obtain $\mathbb{P}(V_t^*XV_t)=\mathbb{P}(U_t^*XU_t)=\mathbb{Q}(X)$.  
Moreover $V_t$ is affiliated to $\mathcal{B}\otimes\mathcal{Z}_t\subset
\mathcal{Z}_t'$, as it is defined by a QSDE which is integrated against
$Z_t$ and has coefficients in $\mathcal{B}$ (this statement can be 
rigorously verified by an approximation argument.) 
Hence by Lemma \ref{lem:bayes} we obtain $\mathbb{Q}(X|\mathcal{Z}_t)=
\mathbb{P}(V_t^*XV_t|\mathcal{Z}_t)/\mathbb{P}(V_t^*V_t|\mathcal{Z}_t)$.
The statement of the Theorem follows immediately. 
\end{proof}

\begin{proof}[Proof of Theorem \ref{thm:KScount}]
The proof proceeds along the same lines as the proof of Theorem 
\ref{thm:KSdiff}, except that in order to apply Lemma \ref{lem:holevo} we 
must make sure that the coefficient in front of $dA_t^*$ in the output 
noise does not vanish.  To this end we perform an additional rotation by 
$E_t$; the expression for $C_t$ is obtained by direct application of the 
quantum It\^o rules.  If we introduce the transformed state $\mathbb{Q}(X)=
\mathbb{P}(U_t^*E_tXE_t^*U_t)$, we obtain $\pi_t(X)=
U_t^*E_t\mathbb{Q}(X|\mathcal{C}_t)E_t^*U_t$ for $X\in\mathcal{B}$ (we 
have used the fact that any such $X$ commutes with $E_t$.)  It remains to 
notice that $\mathbb{P}(V_t^*XV_t)=\mathbb{Q}(X)$ by Lemma 
\ref{lem:holevo} and by application of the quantum It\^o rules to 
$E_t^*U_t$.  The statement of the Theorem follows from Lemma 
\ref{lem:bayes}.
\end{proof}

Now that we have the quantum Kallianpur-Striebel formulas, it is not 
difficult to obtain a recursive representation of the filtering equations.
Let us demonstrate the procedure with Theorem \ref{thm:KSdiff}.  Using the 
quantum It\^o rules, we can write 
\begin{multline*}
	V_t^*XV_t=X+\int_0^t V_s^*(
		i[H_s,X]+L_s^*XL_s-\tfrac{1}{2}\{
		L_s^*L_sX+XL_s^*L_s
	\})V_s\,ds \\
	+\int_0^t
	V_s^*(\alpha_s^{-1}XL_s+\alpha_s^{-1*}L_s^*X
		+(\alpha_s^*\alpha_s)^{-1}\lambda_sL_s^*XL_s
	)V_s\,dZ_t
\end{multline*}
for every $X\in\mathcal{B}$ (and we use the shortened notation
$dZ_t=\lambda_s\,d\Lambda_s+\alpha_s\,dA_s^*+\alpha_s^*\,dA_s$.)
Taking the conditional expectation of both sides, we obtain
\begin{multline*}
	\mathbb{P}(V_t^*XV_t-X|\mathcal{Z}_t)=
	\int_0^t 
	\mathbb{P}(V_s^*(
		i[H_s,X]+L_s^*XL_s-\tfrac{1}{2}\{
		L_s^*L_sX+XL_s^*L_s
	\})V_s|\mathcal{Z}_s)\,ds \\
	+\int_0^t\mathbb{P}(
	V_s^*(\alpha_s^{-1}XL_s+\alpha_s^{-1*}L_s^*X
		+(\alpha_s^*\alpha_s)^{-1}\lambda_sL_s^*XL_s
	)V_s|\mathcal{Z}_s)\,dZ_t,
\end{multline*}
where the fact that we can pull the conditional expectation into the 
integrals can once again be verified by an approximation argument.  
It remains to rotate the expression by $U_t$; another application of the 
quantum It\^o rules gives
\begin{multline*}
	\sigma_t(X)=\mathbb{P}(X)+\int_0^t \sigma_s(
		i[H_s,X]+L_s^*XL_s-\tfrac{1}{2}\{
		L_s^*L_sX+XL_s^*L_s
	\})\,ds \\
	+\int_0^t
	\sigma_s(\alpha_s^{-1}XL_s+\alpha_s^{-1*}L_s^*X
		+(\alpha_s^*\alpha_s)^{-1}\lambda_sL_s^*XL_s
	)\,dY_t
\end{multline*}
where $dY_t$ is given by (\ref{eq:sysobs}).  This is the noncommutative 
counterpart of the linear filtering equation of classical nonlinear 
filtering theory.  Applying a similar procedure to Theorem 
\ref{thm:KScount} yields a linear filtering equation for the counting 
case.

\section{Controlled quantum flows and controlled filtering}
\label{sec:controlqf}

In the previous section we considered a quantum system where the 
coefficients $H_t,L_t,S_t$ were deterministic functions in $\mathcal{B}$;
for example, we considered the case where $L,S$ were constant in time and 
where the Hamiltonian $H_t=u(t)H$ was modulated by a deterministic 
control.  This gives rise to a filtering equation, e.g.\ the linear 
filtering equation that propagates $\sigma_t(\cdot)$, which also depends 
deterministically on the control $u(t)$.  In a feedback control scenario, 
however, we would like to adapt the controls in real time based on the 
observations that have been accumulated; i.e., we want to make $u(t)$ a 
function of the observations $Y_{s}$ up to time $t$.  In this section we 
introduce the notion of a controlled QSDE, and show that this gives rise 
to a controlled linear filtering equation of the same form as in the 
previous section.

\begin{definition}[Controlled quantum flow]
\label{def:cdiff}
Given
\begin{enumerate}
\item an {\em output noise} $Z_t$ of the form
$$
	Z_t=\int_0^t \Xi_s\,d\Lambda_s+
		\int_0^t\Upsilon_s\,dA_s^*+
		\int_0^t\Upsilon_s^*\,dA_s
$$
such that $\Xi_t,\Upsilon_t$ are adapted and affiliated to 
$\mathcal{Z}_t={\rm vN}(Z_s:0\le s\le t)$ for every $t\in[0,T]$, $\Xi_t$ 
is self-adjoint, and $\mathcal{Z}_t$ is a commutative algebra; and
\item a {\em controlled Hudson-Parthasarathy equation}
$$
	U_t=I
	+\int_0^tL_sU_s\,dA_s^*
	-\int_0^tL_s^*S_sU_s\,dA_s 
	+\int_0^t(S_s-I)U_s\,d\Lambda_s
	-\int_0^t(
		iH_s+\tfrac{1}{2}L_s^*L_s
	)U_s\,ds
$$
where $H_t,L_t,S_t$ are affiliated to $\mathcal{B}\otimes\mathcal{Z}_t$
for every $t\in[0,T]$,
\end{enumerate}
the pair $(j_t,Y_t)$, where $j_t(X)=U_t^*XU_t$ $(X\in 
\mathcal{B}\otimes\mathcal{Z}_t)$ and $Y_t=U_t^*Z_tU_t$, 
is called a {\em controlled quantum flow} $j_t$ with {\em observation 
process} $Y_t$.
\end{definition}

To use this definition we must impose sufficient regularity conditions on 
the various processes involved so that these are indeed well defined.  
From this point onward we assume that $\Xi_t,\Upsilon_t,H_t,L_t,S_t$ 
are bounded measurable processes, i.e.\ 
$$
	\sup_{t\in[0,T]}\|\Xi_t\|<\infty,\qquad
	t\mapsto\Xi_t\psi \mbox{ is measurable }
	\forall\,\psi\in\mathscr{H}_0\otimes\Gamma,
$$
and similarly for the other processes.  Under such assumptions 
(essentially bounded control requirements) we can show \cite{BvH05b} that 
$Z_t$ is well defined and that $U_t$ has a unique unitary solution (see also
\cite{Mey93,Hol96,GLW01} for related results.)

Before we discuss filtering in the context of Definition \ref{def:cdiff},
let us clarify the significance of this definition.  First, note that we 
can use the quantum It\^o rules to obtain the system-observations pair
\begin{multline}\label{eq:csyssys}
	dj_t(X)=
	j_t(i[H_t,X]
		+L_t^*XL_t-\tfrac{1}{2}\{L_t^*L_tX+
		XL_t^*L_t\})\,dt \\
	+j_t(S_t^*[X,L_t])\,dA_t^*
	+j_t([L_t^*,X]S_t)\,dA_t
	+j_t(S_t^*XS_t-X)\,d\Lambda_t,
\end{multline}
\begin{multline}\label{eq:csysobs}
	dY_t = \Xi_t\,d\Lambda_t
	+j_t\left(
		S_t^*(\Upsilon_t+\Xi_tL_t)
	\right)\,dA_t^*
	+j_t\left(
		(\Upsilon_t^*+\Xi_tL_t^*)S_t
	\right)\,dA_t \\
	+j_t\left(
		\Xi_tL_t^*L_t+\Upsilon_t^*L_t+\Upsilon_tL_t^*
	\right)\,dt,
\end{multline}
which is simply the controlled counterpart of (\ref{eq:syssys}),
(\ref{eq:sysobs}).  The essential thing to notice is that though the
quantities $H_t,L_t$ etc.\ that appear in the equation for the flow $U_t$
are affiliated to the output noise $\mathcal{Z}_t$, the quantities that
appear in the system-observations model are in fact of the form
$j_t(H_t)$, etc., which are affiliated to the observations
$\mathcal{Y}_t$. Our model is extremely general and allows for any of the
coefficients of the QSDE, and even the measurement performed in the field,
to be adapted in real time based on the observed process.  To illustrate
the various types of control that are typically used, we give the
following examples.

\begin{example}[Hamiltonian feedback]\label{ex:hamiltonian}
Consider the controlled quantum flow 
$$
	dU_t=(L\,dA_t^*-L^*\,dA_t-\tfrac{1}{2}L^*L\,dt
	-iu_t(Z_{s\le t})H\,dt)\,U_t,\qquad
	Z_t=A_t+A_t^*.
$$
That is, we have chosen $S_t=0$, fixed $L,H\in\mathcal{B}$, $H=H^*$, and 
$u_t(Z_{s\le t})$ is a bounded (real) scalar function of the
output noise up to time $t$.  This gives the system-observation pair
\begin{multline*}
	dj_t(X)=j_t([X,L])\,dA_t^*+j_t([L^*,X])\,dA_t \\
		+j_t(L^*XL-\tfrac{1}{2}(L^*LX+XL^*L))\,dt
		+u_t(Y_{s\le t})\,j_t(i[H,X])\,dt
\end{multline*}
and $dY_t=dA_t+dA_t^*+j_t(L+L^*)\,dt$, where we have pulled the 
control outside $j_t$.  This scenario corresponds to a fixed system-probe 
interaction, measurement and system Hamiltonian, where we allow ourselves 
to feed back some function of the observation history to modulate the 
strength of the Hamiltonian; see e.g.\ \cite{HSM05}.
\end{example}

\begin{example}[Coherent feedback]
The controlled quantum flow defined by
\begin{multline*}
	dU_t=((L+u_t(Z_{s\le t})I)\,dA_t^*
		-(L^*+u_t(Z_{s\le t})^*I)\,dA_t \\
		-\tfrac{1}{2}(L^*L+u_t(Z_{s\le t})^*u_t(Z_{s\le t})I
			+2u_t(Z_{s\le t})L^*)\,dt
	)\,U_t,
\end{multline*}
where $Z_t=A_t+A_t^*$, describes an initial system driven by a field in a 
coherent state, where we modulate the coherent state amplitude through the 
bounded (complex) control $u_t(Z_{s\le t})$ \cite{Bou04b}.  As in the 
previous example, the control becomes a function of the observations
when we transform to the system-theoretic description.
\end{example}

\begin{example}[Adaptive measurement]  In this scenario we choose 
an uncontrolled Hudson-Parthasarathy equation 
$dU_t=(L\,dA_t^*-L^*\,dA_t-\tfrac{1}{2}L^*L\,dt-iH\,dt)\,U_t$, but the 
measurement in the probe field is adapted in real time by
$$
	dZ_t=e^{-iu_t(Z_{s\le t})}dA_t^*+e^{iu_t(Z_{s\le t})}dA_t
$$
where $u_t(Z_{s\le t})$ is a real scalar control function.  This gives 
rise to the observations
$$
	dY_t=dZ_t+\left[j_t(L)\,e^{iu_t(Y_{s\le t})}
	+j_t(L^*)\,e^{-iu_t(Y_{s\le t})}\right]\,dt.
$$
Evidently the control determines which of the system observables
$L\,e^{iu}+L^*\,e^{-iu}$ is detected in the probe.  The possibility to
adapt the measurement in real time is useful for the detection of
quantities that are not described by a system observable, such as the
phase of an optical pulse.  See e.g.\ \cite{Wi95,AASDM02}. 
\end{example}

We now wish to solve the filtering problem for a controlled quantum flow.  
It turns out that when we use the reference probability method, very 
little changes in the procedure outlined above.  In particular, the proofs 
of Proposition \ref{pro:nondemolition} and Lemma \ref{lem:holevo} extend 
readily to the controlled case, and it is straightforward to extend the 
proofs of Theorems \ref{thm:KSdiff} and \ref{thm:KScount} to prove the 
following statements.

\begin{theorem}[Kallianpur-Striebel formula, diffusive case]\label{thm:CKSdiff}
Suppose that $\Upsilon_t$ has a bounded inverse for all $t\in[0,T]$.  Let 
$V_t$ be the solution of the QSDE
$$
	V_t=I+\int_0^t
		(\Xi_s\,d\Lambda_s+\Upsilon_s\,dA_s^*+\Upsilon_s^*\,dA_s)\,
		\Upsilon_s^{-1}L_sV_s
	-\int_0^t(iH_s+\tfrac{1}{2}L_s^*L_s)V_s\,ds.
$$
Then $V_t$ is affiliated to $\mathcal{Z}_t'$ for every $t\in[0,T]$, and
we have the representation $\pi_t(X)=\sigma_t(X)/\sigma_t(I)$ with
$\sigma_t(X)=U_t^*\mathbb{P}(V_t^*XV_t|\mathcal{Z}_t)U_t$,
$X\in\mathcal{B}\otimes\mathcal{Z}_t$.
\end{theorem}

\begin{theorem}[Kallianpur-Striebel formula, counting case]\label{thm:CKScount}
Suppose that $\Xi_t$ has a bounded inverse for all $t\in[0,T]$.  Let $E_t$ 
be the solution of the QSDE
$$
	E_t=I+
	\int_0^t\Xi_s^{-1}(1-\Upsilon_s)E_s\,dA_s^*-
	\int_0^t\Xi_s^{-1}(1-\Upsilon_s^*)E_s\,dA_s-
	\frac{1}{2}\int_0^t
		\Xi_s^{-2}|1-\Upsilon_s|^2E_s\,ds
$$
where $|X|^2=X^*X$.  Let $C_t=E_t^*Z_tE_t$, $\mathcal{C}_t=
{\rm vN}(C_s:0\le s\le t)=E_t^*\mathcal{Z}_tE_t$, so that
$$
	C_t=A_t^*+A_t+\int_0^tE_s^*\Xi_sE_s\,d\Lambda_s+
	\int_0^tE_s^*\Xi_s^{-1}(1-\Upsilon_s^*\Upsilon_s)E_s\,ds.
$$
Define $V_t$ as the solution of the QSDE
\begin{multline*}
	V_t=I+\int_0^t
		(E_s^*\Xi_sE_s\,d\Lambda_s+dA_s^*+dA_s) \,
		E_s^*(L_s-\Xi_s^{-1}(1-\Upsilon_s))E_sV_s \\
	-\int_0^tE_s^*(iH_s+\tfrac{1}{2}L_s^*L_s
		+\tfrac{1}{2}\Xi_s^{-2}|1-\Upsilon_s|^2
		-\Xi_s^{-1}(1-\Upsilon_s^*)L_s)E_sV_s\,ds.
\end{multline*}
Then $V_t$ is affiliated to $\mathcal{C}_t'$ for every $t\in[0,T]$, and we 
have the representation $\pi_t(X)=\sigma_t(X)/\sigma_t(I)$ with
$\sigma_t(X)=U_t^*E_t\mathbb{P}(V_t^*E_t^*XE_tV_t|\mathcal{C}_t)E_t^*U_t$,
$X\in\mathcal{B}\otimes\mathcal{Z}_t$.
\end{theorem}

We can now obtain controlled filtering equations for $\sigma_t(\cdot)$ in 
a recursive form.  The following statements are readily verified using the 
quantum It\^o rules.

\begin{corollary}[linear filtering equation, diffusive case]\label{cor:zakdiff}
Suppose that $\Upsilon_t$ has a bounded inverse for all $t\in[0,T]$.
Then $\sigma_t(\cdot)$ of Theorem \ref{thm:CKSdiff} satisfies
\begin{multline*}
	\sigma_t(X)=\mathbb{P}(X)+\int_0^t \sigma_s(
		i[H_s,X]+L_s^*XL_s-\tfrac{1}{2}\{
		L_s^*L_sX+XL_s^*L_s
	\})\,ds \\
	+\int_0^t
	\sigma_s(\Upsilon_s^{-1}XL_s+\Upsilon_s^{-1*}L_s^*X
		+(\Upsilon_s^*\Upsilon_s)^{-1}\Xi_s\,L_s^*XL_s
	)\,dY_s
\end{multline*}
for $X\in\mathcal{B}$, where $dY_t$ is given by (\ref{eq:csysobs}) and 
$\pi_t(X)=\sigma_t(X)/\sigma_t(I)$.
\end{corollary}

\begin{corollary}[linear filtering equation, counting case]\label{cor:zakcount}
Suppose that $\Xi_t$ has a bounded inverse for all $t\in[0,T]$.
Then $\sigma_t(\cdot)$ of Theorem \ref{thm:CKScount} satisfies
\begin{multline*}
	\sigma_t(X)=\mathbb{P}(X)+\int_0^t \sigma_s(
		i[H_s,X]+L_s^*XL_s-\tfrac{1}{2}\{
		L_s^*L_sX+XL_s^*L_s
	\})\,ds \\
	+\int_0^t
	\sigma_s(
		\Xi_s\,L_s^*XL_s+\Upsilon_s^*XL_s+
		\Upsilon_sL_s^*X-\Xi_s^{-1}(1-\Upsilon_s^*\Upsilon_s)X
	)
		\times \\
	(dY_s-j_s(\Xi_s^{-1}(1-\Upsilon_s^*\Upsilon_s))\,ds)
\end{multline*}
for $X\in\mathcal{B}$, where $dY_t$ is given by (\ref{eq:csysobs}) and 
$\pi_t(X)=\sigma_t(X)/\sigma_t(I)$.
\end{corollary}

What is the relation between the open loop filters of the previous section
and the closed loop filters associated to a controlled quantum flow?
Consider for example the case of Hamiltonian control where
$Z_t=A_t+A_t^*$, $S_t=0$, $L_t=L\in\mathcal{B}$ is constant and
$H_t=u(t)H$ with $H\in\mathcal{B}$. In open loop $u(t)$ is a deterministic
function and we saw in the previous section that this gives rise to the
linear filtering equation
$$
	d\sigma_t(X)=u(t)\,\sigma_t(i[H,X])\,dt+
	\sigma_t(L^*XL-\tfrac{1}{2}\{L^*LX+XL^*L\})\,dt
	+\sigma_t(XL+L^*X)\,dY_t.
$$
In closed loop the model is given by the controlled quantum flow of 
Example \ref{ex:hamiltonian}, and by Corollary \ref{cor:zakdiff} we obtain 
the controlled linear filtering equation
\begin{multline*}
	d\sigma_t(X)=u(Y_{s\le t})\,\sigma_t(i[H,X])\,dt \\
	+\sigma_t(L^*XL-\tfrac{1}{2}\{L^*LX+XL^*L\})\,dt
	+\sigma_t(XL+L^*X)\,dY_t.
\end{multline*}
Evidently we obtain the same filter for the closed loop controlled quantum
flow as we would obtain by calculating the open loop filter and then
substituting a feedback control for the deterministic function $u(t)$.
This is in fact a general property of controlled filtering equations, as
can be seen directly from the statement of Corollaries \ref{cor:zakdiff}
and \ref{cor:zakcount}.  Though this property is usually assumed to hold
true in the literature, we see here that it follows from the definition of 
a controlled quantum flow.

\section{Sample path properties and the innovations problem}
\label{sec:innovations}

In the remainder of this paper we will use explicitly the properties of
quantum filtering equations in recursive form, as given e.g.\ in
Corollaries \ref{cor:zakdiff} and \ref{cor:zakcount}.  In the next section
we will show that under suitable regularity conditions, the quantum
optimal control problem is solved by a feedback control policy that at
time $t$ is only a function of the normalized solution of the linear 
filtering equation at that time.  In order for this to be sensible, we 
have to show that the controlled linear filtering equation, and in 
particular its normalized form, has a unique strong solution.  The purpose 
of this section is to investigate these and related properties of the 
solutions of recursive quantum filters.

The approach we will take is to convert the entire problem into one of
classical stochastic analysis.  Note that all the quantities that appear
in the linear filtering equations of Corollaries \ref{cor:zakdiff} and
\ref{cor:zakcount} are adapted and affiliated to the commutative algebra
$\mathcal{Y}_t$.  Thus, we may use the spectral theorem to map the filter
onto a classical stochastic differential equation driven by the
(classical) observations.  This will allow us to manipulate the filter by
using the It\^o change of variables formula for jump-diffusions and puts
at our disposal the full machinery of classical stochastic differential
equations driven by semimartingales.

We begin by proving the following Proposition.  This property will be 
crucial in the proof of the separation theorem; at this point, however, we 
are mostly interested in the fact that as a consequence, the observation
process $Y_t$ is a semimartingale.

\begin{proposition}[Innovations martingale]\label{pro:innovation}
Let $dZ_t=\Xi_t\,d\Lambda_t+\Upsilon_t\,dA^*_t+\Upsilon^*_t\,dA_t$ 
as in Definition \ref{def:cdiff}.  Define the innovations process
$$
	\overline{Z}_t = U_t^*Z_tU_t - \int_0^t\pi_s(
		\Xi_sL_s^*L_s 
		+\Upsilon_s^*L_s
		+\Upsilon_sL^*_s
	)\,ds.
$$
Then $\overline{Z}_t$ is a $\mathcal{Y}_t$-martingale, i.e.\ for all 
$s\le t\in[0,T]$ we have $\mathbb{P}(\overline{Z}_t|\mathcal{Y}_s) = 
\overline{Z}_s$.
\end{proposition}

\begin{proof} 
We need to prove that
$\mathbb{P}(\overline{Z}_t-\overline{Z}_s|\mathcal{Y}_s)=0$ for all $s\le
t\in[0,T]$, or equivalently $\mathbb{P}((\overline{Z}_t-\overline{Z}_s)K)=0$
for all $s\le t\in[0,T]$ and $K\in\mathcal{Y}_s$.  The latter can be 
written as
$$
	\mathbb{P}(U_t^*Z_tU_tK) - \mathbb{P}(U^*_sZ_sU_sK) = 
	\int_s^t \mathbb{P}(\pi_s(
		\Xi_sL_s^*L_s
		+\Upsilon_s^*L_s
		+\Upsilon_sL^*_s
	)K)\,ds
$$
for all $s\le t\in[0,T]$ and $K\in\mathcal{Y}_s$.  Since 
$\mathcal{Y}_s=U_t^*\mathcal{Z}_sU_t$ for all $s\le t\in[0,T]$, it is 
sufficient to show that
$$
	\mathbb{P}(U_t^*Z_tCU_t) - \mathbb{P}(U^*_sZ_sCU_s) = 
	\int_s^t \mathbb{P}(U_t^*(
		\Xi_sL_s^*L_sC
		+\Upsilon_s^*L_sC
		+\Upsilon_sL^*_sC
	)U_t)\,ds
$$
for all $s\le t\in[0,T]$ and $K\in\mathcal{Z}_s$.  But this follows 
directly from (\ref{eq:csysobs}).
\end{proof}

The innovations process is the starting point for martingale-based
approaches to (quantum) filtering \cite{Bel92b,BGM04}.  The idea there is
to obtain a particular martingale which is represented as a stochastic
integral with respect to the innovations.  The method is complicated, 
however, by what is known as the innovations problem: it is not clear a
priori whether the observations and the innovations generate the same
($\sigma$-)algebras \cite{LiS01}, which is a prerequisite for the
martingale representation theorem.  The problem is resolved using a method
by Fujisaki-Kallianpur-Kunita, where the Girsanov theorem is used to prove
a special martingale representation theorem with respect to the
innovations \cite{LiS01}.  In contrast, the reference probability method
is completely independent from the innovations problem. Though we will not
need this fact to prove the separation theorem, we will see at the end of
this section that the equivalence of the observations and innovations
$\sigma$-algebras follows as a corollary from the existence and uniqueness
theorems.

To return to the task at hand, it is evident from Proposition
\ref{pro:innovation} that the observation process $Y_t$ can be written as
the sum of the innovations process, which is a martingale, and a process
of finite variation, both of which are affiliated to the algebra generated
by the observations.  Hence if we map $Y_t$ to its classical counterpart
through the spectral theorem, we obtain a classical semimartingale.  

\begin{remark} 
For convenience, we will abuse our notation somewhat and denote by
$Y_t,\overline{Z}_t,\pi_t(X),\sigma_t(X)$ both the corresponding quantum 
processes and the associated classical processes obtained through the 
spectral theorem. By $\tilde\Xi_t,\tilde\Upsilon_t$ we denote the 
classical processes obtained by applying the spectral theorem to 
$j_t(\Xi_t)$ and $j_t(\Upsilon_t)$, whereas $\tilde L_t,\tilde H_t$ are 
$(\dim\mathscr{H}_0\times\dim\mathscr{H}_0)$-matrix valued processes 
obtained by applying the spectral theorem to each matrix element of 
$j_t(L_t)$ and $j_t(H_t)$.  Hence $\tilde\Xi_t,\tilde\Upsilon_t$ are 
$Y_t$-adapted bounded scalar processes, whereas $j_t(L_t),j_t(H_t)$ are 
$Y_t$-adapted bounded matrix-valued processes.
\end{remark}

We now map the linear filtering equation of Corollary \ref{cor:zakdiff} to 
a classical stochastic differential equation.  This gives
\begin{multline*}
	\sigma_t(X)=
	\mathbb{P}(X)+\int_0^t \sigma_{s}(
		i[H_{s},X]+L_{s}^*XL_{s}-\tfrac{1}{2}\{
		L_{s}^*L_{s}X+XL_{s}^*L_{s}
	\})\,ds \\
	+\int_0^t
	\sigma_{s-}(\Upsilon_{s-}^{-1}XL_{s-}+\Upsilon_{s-}^{-1*}L_{s-}^*X
		+(\Upsilon_{s-}^*\Upsilon_{s-})^{-1}\Xi_{s-}\,L_{s-}^*XL_{s-}
	)\,dY_s
\end{multline*}
and similarly, we obtain the classical equivalent of Corollary 
\ref{cor:zakcount}
\begin{multline*}
	\sigma_t(X)=
	\mathbb{P}(X)+\int_0^t \sigma_{s}(
		i[H_{s},X]+L_{s}^*XL_{s}-\tfrac{1}{2}\{
		L_{s}^*L_{s}X+XL_{s}^*L_{s}
	\})\,ds \\
	+\int_0^t
	\sigma_{s-}(
		\Xi_{s-}\,L_{s-}^*XL_{s-}+\Upsilon_{s-}^*XL_{s-}+
		\Upsilon_{s-}L_{s-}^*X
		-\Xi_{s-}^{-1}(1-\Upsilon_{s-}^*\Upsilon_{s-})X
	)
		\times \\
	(dY_s-
	\tilde\Xi_{s}^{-1}(1-\tilde\Upsilon_{s}^*\tilde\Upsilon_{s})\,ds)
\end{multline*}
where now the stochastic integrals are classical It\^o integrals with
respect to the semimartingale $Y_t$ \cite{Pro04} (the fact that we can map
a quantum It\^o integral with respect to fundamental processes to a
classical It\^o integral with respect to a semimartingale is verified by
approximation.)  As we are now dealing with stochastic processes on the 
level of sample paths, we have to choose a modification such that the 
processes are well defined---this is an issue that does not occur on the 
level of QSDEs.  We will make the standard choice \cite{Pro04} that all 
our (semi)martingales are c{\`a}dl{\`a}g, and include explicitly the left 
limits $\sigma_{s-}$ etc.\ to enforce causality.

We are now ready to apply classical stochastic analysis to our quantum 
filtering equations.  We begin by normalizing the equations using the 
classical It\^o formula.

\begin{proposition}[nonlinear filtering equation, pure diffusion case]
\label{pro:KSdiff}
Suppose that $\Xi_t=0$ and $\Upsilon_t$ has a bounded inverse for all 
$t\in[0,T]$.  Then $\pi_t(\cdot)$ satisfies with respect to the 
semimartingale observations $Y_t$ the It\^o equation
\begin{multline*}
	\pi_t(X)=\mathbb{P}(X)+\int_0^t \pi_s(
		i[H_s,X]+L_s^*XL_s-\tfrac{1}{2}\{
		L_s^*L_sX+XL_s^*L_s
	\})\,ds \\
	+\int_0^t\left\{
	\pi_{s}(\Upsilon_{s}^{-1}XL_{s}+\Upsilon_{s}^{-1*}L_{s}^*X)
	-\pi_{s}
	(\Upsilon_{s}^{-1}L_{s}+\Upsilon_{s}^{-1*}L_{s}^*)\,\pi_{s}(X)
	\right\}d\overline{Z}_s
\end{multline*}
for $X\in\mathcal{B}$, where $\overline{Z}_t$ is the innovations process
of Proposition \ref{pro:innovation}.  Furthermore, there is a
$(\dim\mathscr{H}_0\times\dim\mathscr{H}_0)$-matrix process $\rho_t$ such
that $\pi_t(X)={\rm Tr}[X\rho_t]$ for all $X\in\mathcal{B}$, which
satisfies the classical It\^o stochastic differential equation
\begin{multline*}
	\rho_t=\rho_0+\int_0^t\left\{
		-i[\tilde H_s,\rho_s]
		+\tilde L_s\rho_s\tilde L_s^*-\tfrac{1}{2}(
		\tilde L_s^*\tilde L_s\rho_s+\rho_s\tilde L_s^*\tilde L_s
	)\right\}ds \\
	+\int_0^t\left\{
	\tilde\Upsilon_{s}^{-1}\tilde L_{s}\rho_{s}
	+\tilde\Upsilon_{s}^{-1*}\rho_{s}\tilde L_{s}^*
	-{\rm Tr}[(\tilde\Upsilon_{s}^{-1}\tilde L_{s}
		+\tilde\Upsilon_{s}^{-1*}\tilde L_{s}^*)\rho_{s}]\,\rho_{s}
	\right\}d\overline{Z}_s.
\end{multline*}
\end{proposition}

\begin{proposition}[nonlinear filtering equation, pure jump case]
\label{pro:KSjump}
Suppose that $\Xi_t$ has a bounded inverse for all $t\in[0,T]$.  Then 
$\pi_t(\cdot)$ satisfies with respect to the semimartingale observations 
$Y_t$ the It\^o equation
\begin{multline*}
	\pi_t(X)=
	\mathbb{P}(X)+\int_0^t \pi_{s}(
		i[H_{s},X]+L_{s}^*XL_{s}-\tfrac{1}{2}\{
		L_{s}^*L_{s}X+XL_{s}^*L_{s}
	\})\,ds \\
	+\int_0^t\left\{
		\frac{\pi_{s-}((\Upsilon_{s-}+\Xi_{s-}L_{s-})^*X
			(\Upsilon_{s-}+\Xi_{s-}L_{s-}))}{
		\pi_{s-}((\Upsilon_{s-}+\Xi_{s-}L_{s-})^*
			(\Upsilon_{s-}+\Xi_{s-}L_{s-}))}
	-\pi_{s-}(X)
	\right\}\tilde\Xi_{s-}^{-1}\,d\overline{Z}_s
\end{multline*}
for $X\in\mathcal{B}$, where $\overline{Z}_t$ is the innovations process
of Proposition \ref{pro:innovation}. Furthermore, there is a
$(\dim\mathscr{H}_0\times\dim\mathscr{H}_0)$-matrix process $\rho_t$ such
that $\pi_t(X)={\rm Tr}[X\rho_t]$ for all $X\in\mathcal{B}$, which
satisfies the classical It\^o stochastic differential equation
\begin{multline*}
	\rho_t=\rho_0+\int_0^t\left\{
		-i[\tilde H_s,\rho_s]
		+\tilde L_s\rho_s\tilde L_s^*-\tfrac{1}{2}(
		\tilde L_s^*\tilde L_s\rho_s+\rho_s\tilde L_s^*\tilde L_s
	)\right\}ds \\
	+\int_0^t\left\{
	\frac{
		(\tilde\Upsilon_{s-}+\tilde\Xi_{s-}\tilde L_{s-})\rho_{s-}
			(\tilde\Upsilon_{s-}+\tilde\Xi_{s-}\tilde L_{s-})^*
	}{
		{\rm Tr}[(\tilde\Upsilon_{s-}
			+\tilde\Xi_{s-}\tilde L_{s-})\rho_{s-}
			(\tilde\Upsilon_{s-}+\tilde\Xi_{s-}\tilde L_{s-})^*]
	}
	-\rho_{s-}
	\right\}\tilde\Xi_{s-}^{-1}\,d\overline{Z}_s.
\end{multline*}
\end{proposition}

\begin{proof}[Proof of Propositions \ref{pro:KSdiff} and \ref{pro:KSjump}]
In pure diffusion case (Proposition \ref{pro:KSdiff}) the observation
process $Y_t$ is a continuous semimartingale, and hence the normalization
is easily verified by applying the It\^o change of variables formula to
the Kallianpur-Striebel formula $\pi_t(X)=\sigma_t(X)/\sigma_t(I)$.  In
both cases, the conversion to the matrix form follows from the fact that
by construction $\pi_t(X)$ is a random linear functional on $\mathcal{B}$,
and hence there is a unique random matrix $\rho_t$ such that 
$\pi_t(X)={\rm Tr}[X\rho_t]$ for all $X\in\mathcal{B}$.  The expressions 
given are easily seen to satisfy this requirement.

It remains to normalize $\sigma_t(X)$ of Proposition \ref{pro:KSjump}.  
Here we use the It\^o change of variables formula for Stieltjes integrals, 
but the manipulations are somewhat more cumbersome than in the continuous 
case.  We begin by noting that
$$
	\tilde Y_t=\int_0^t\tilde\Xi_{s-}^{-1}(dY_s+
		\tilde\Xi_{s}^{-1}
		\tilde\Upsilon_s^*\tilde\Upsilon_s\,ds)
$$
is a pure jump process with jumps of unit magnitude.  To see this, we 
calculate
$$
	[\tilde Y,\tilde Y]_t=\tilde Y_t^2
		-2\int_0^t\tilde Y_{s-}\,d\tilde Y_{s}=
	\tilde Y_t
$$
by applying the quantum It\^o rules to (\ref{eq:csysobs}) and using the 
spectral theorem.  The quadratic variation $[\tilde Y,\tilde Y]_t$ is by 
construction an increasing process, so $\tilde Y_t$ is also increasing and 
hence of finite variation. But any adapted, c{\`a}dl{\`a}g finite 
variation process is a quadratic pure jump semimartingale \cite{Pro04}, 
meaning that
$$
	\tilde Y_t=[\tilde Y,\tilde Y]_t=
	\sum_{0<s\le t}(\tilde Y_s-\tilde Y_{s-})^2.
$$
As $\tilde Y_t$ is an increasing pure jump process and $\tilde Y_s-\tilde 
Y_{s-}=(\tilde Y_s-\tilde Y_{s-})^2$, we conclude that $\tilde Y_t$ is a 
pure jump process with unit magnitude jumps.
We can now rewrite the linear filtering equation for the pure jump case in 
terms of $\tilde Y_t$.  This gives
\begin{multline*}
	\sigma_t(X)=
	\mathbb{P}(X)+\int_0^t \sigma_{s}(
		i[H_{s},X]+L_{s}^*XL_{s}-\tfrac{1}{2}\{
		L_{s}^*L_{s}X+XL_{s}^*L_{s}
	\})\,ds \\
	+\int_0^t
	\sigma_{s-}(
		(\Upsilon_{s-}+\Xi_{s-}L_{s-})^*X
		(\Upsilon_{s-}+\Xi_{s-}L_{s-})-X
	)\,(d\tilde Y_s-\tilde\Xi_{s}^{-2}\,ds)
\end{multline*}
where the integral over $\tilde Y_t$ is a simple Stieltjes integral.  To 
normalize $\sigma_t(X)$, we first use the change of variables formula for 
finite variation processes \cite{Pro04}
$$
	f(V_t)-f(V_0)=\int_0^t f'(V_{s-})\,dV_s
	+\sum_{0<s\le t}\left\{
		f(V_s)-f(V_{s-})-f'(V_{s-})\,(V_s-V_{s-})
	\right\}
$$
to calculate $\sigma_t(I)^{-1}$.  We obtain directly that
$$
	\sigma_s(I)-\sigma_{s-}(I)=
	(\sigma_{s-}(B_{s-}^*B_{s-})-\sigma_{s-}(I))\,
	(\tilde Y_s-\tilde Y_{s-})	
$$
where $B_{s}=\Upsilon_{s}+\Xi_{s}L_{s}$.  Similarly we have
$$
	\sigma_s(I)^{-1}-\sigma_{s-}(I)^{-1}=
	(\sigma_{s-}(B_{s-}^*B_{s-})^{-1}-\sigma_{s-}(I)^{-1})\,
	(\tilde Y_s-\tilde Y_{s-}).
$$
We can thus express the correction term in the change of variables formula 
as
\begin{multline*}
	\sum_{0<s\le t}\left\{
		\sigma_s(I)^{-1}
		-\sigma_{s-}(I)^{-1}
		+\sigma_{s-}(I)^{-2}\,(\sigma_s(I)-\sigma_{s-}(I))
	\right\}= \\
	\int_0^t
		(\pi_{s-}(B_{s-}^*B_{s-})+\pi_{s-}(B_{s-}^*B_{s-})^{-1}-2)
	\,\sigma_{s-}(I)^{-1}\,
	d\tilde Y_s
\end{multline*}
and we obtain
$$
	\sigma_t(I)^{-1}=1+\int_0^t
	\frac{\pi_s(B_s^*B_s)-1}{\sigma_s(I)}\,\tilde\Xi_s^{-2}\,ds
	+\int_0^t\frac{\pi_{s-}(B_{s-}^*B_{s-})^{-1}-1}{\sigma_{s-}(I)}
		\,d\tilde Y_s.
$$
Finally, applying the integration by parts formula for Stieltjes 
integrals to the Kallianpur-Striebel formula 
$\pi_t(X)=\sigma_t(X)/\sigma_t(I)$ gives
\begin{multline*}
	\pi_t(X)=
	\mathbb{P}(X)+\int_0^t \pi_{s}(
		i[H_{s},X]+L_{s}^*XL_{s}-\tfrac{1}{2}\{
		L_{s}^*L_{s}X+XL_{s}^*L_{s}
	\})\,ds \\
	+\int_0^t\left\{
		\frac{\pi_{s-}(B_{s-}^*XB_{s-})}{\pi_{s-}(B_{s-}^*B_{s-})}
		-\pi_{s-}(X)
	\right\}\,(d\tilde Y_s-\tilde\Xi_s^{-2}\,\pi_s(B_s^*B_s)\,ds).
\end{multline*}
and it remains to notice that $\tilde\Xi_{s-}^{-1}\,d\overline{Z}_s=
d\tilde Y_s-\tilde\Xi_s^{-2}\,\pi_s(B_s^*B_s)\,ds$.
\end{proof}

Now that we have obtained the filtering equations in their sample path 
form, let us study the question of existence and uniqueness of solutions.

\begin{proposition}[Existence and uniqueness]
Let $\tilde\Xi_s$, $\tilde\Upsilon_s$ and the matrix elements of 
$\tilde L_s$ and $\tilde H_s$ have c{\`a}dl{\`a}g sample paths.  Then both 
the linear and nonlinear filtering equations have a unique strong solution 
with respect to $Y_t$.  Moreover, matrix $\rho_t$ is a.s.\ positive with 
unit trace, i.e.\ a density matrix, for every $t$.
\end{proposition}

\begin{proof}
As $\sigma_t(X)$ is linear by construction, we can find a matrix process 
$\tau_t$ such that $\sigma_t(X)={\rm Tr}[X\tau_t]$.  As we did for 
$\rho_t$, we obtain directly that $\tau_t$ must satisfy
\begin{multline*}
	\tau_t=\rho_0+\int_0^t\left\{
		-i[\tilde H_s,\tau_s]
		+\tilde L_s\tau_s\tilde L_s^*-\tfrac{1}{2}(
		\tilde L_s^*\tilde L_s\tau_s+\tau_s\tilde L_s^*\tilde L_s
	)\right\}ds \\
	+\int_0^t\left\{
	\tilde\Upsilon_s^{-1}\tilde L_s\tau_s
	+\tilde\Upsilon_s^{-1*}\tau_s\tilde L_s^*
	\right\}dY_s,
\end{multline*}
in the pure diffusion case, and in the pure jump case $\tau_t$ must 
satisfy
\begin{multline*}
	\tau_t=\rho_0+\int_0^t\left\{
		-i[\tilde H_s,\tau_s]
		+\tilde L_s\tau_s\tilde L_s^*-\tfrac{1}{2}(
		\tilde L_s^*\tilde L_s\tau_s+\tau_s\tilde L_s^*\tilde L_s
	)\right\}ds \\
	+\int_0^t\left\{
	(\tilde\Upsilon_{s-}+\tilde\Xi_{s-}\tilde L_{s-})\tau_{s-}
	(\tilde\Upsilon_{s-}+\tilde\Xi_{s-}\tilde L_{s-})^*-\tau_{s-}
	\right\}\times\\
	\tilde\Xi_{s-}^{-1}(dY_s-
	\tilde\Xi_{s}^{-1}(1-\tilde\Upsilon_{s}^*\tilde\Upsilon_{s})\,ds).
\end{multline*}
Both these equations are finite-dimensional linear stochastic differential
equations with c{\`a}dl{\`a}g coefficients, for which the existence of a
unique strong solution is a standard result \cite{Pro04}.  Thus the unique
solution $\tau_t$ of these equations must indeed satisfy $\sigma_t(X)={\rm
Tr}[X\tau_t]$ (this need not be true if the solution were not 
unique---then it could be the case that only one of the solutions
coincides with $\sigma_t(\cdot)$.)

To demonstrate the existence of a solution to the equations for $\rho_t$, 
note that by construction $\sigma_t(X)$ is a positive map (we will always 
assume that $\rho_0$ is chosen so that $\sigma_0(X)=\pi_0(X)={\rm 
Tr}[X\rho_0]$ is a state, i.e.\ $\rho_0$ is a positive matrix with unit 
trace.)  Hence by uniqueness $\tau_t$ must be a positive matrix for all 
$t$.  Moreover, as the linear solution map $\tau_0\mapsto\tau_t$ is a.s.\ 
invertible \cite{Pro04} $\tau_t$ is for all $t$ a.s.\ not the zero matrix.
This means that for all $t$ the process $\rho_t=\tau_t/{\rm Tr}[\tau_t]$ 
is well-defined and satisfies the nonlinear filtering equations for 
$\rho_t$ which we obtained previously.  Hence we have explicitly 
constructed a solution to the equations for $\rho_t$, and moreover 
$\rho_t$ is a.s.\ a positive, unit trace matrix for every $t$.  Finally, 
from the Kallianpur-Striebel formula it is evident that $\pi_t(X)={\rm 
Tr}[X\rho_t]$.

It remains to prove that there are no other solutions $\rho_t$ that 
satisfy the nonlinear filtering equations, i.e.\ that the solution
$\rho_t$ constructed above is unique.  To this end, suppose that there is 
a different solution $\bar\rho_t$ with $\bar\rho_0=\rho_0$ that also 
satisfies the nonlinear filtering equation (with respect to $dY_t$).  
Define the process $\bar\Pi_t$ as the unique strong solution of the linear 
equation
$$
	\bar\Pi_t=1+\int_0^t{\rm Tr}[(
		\tilde\Upsilon_s^{-1}\tilde L_s
		+\tilde\Upsilon_s^{-1*}\tilde L_s^*
	)\bar\rho_s]\bar\Pi_s\,dY_s
$$
in the pure diffusion case, and as the unique strong solution of
\begin{multline*}
	\bar\Pi_t=1+\int_0^t\left\{{\rm Tr}[
	(\tilde\Upsilon_{s-}+\tilde\Xi_{s-}\tilde L_{s-})^*
	(\tilde\Upsilon_{s-}+\tilde\Xi_{s-}\tilde L_{s-})\bar\rho_{s-}
	]-1\right\}\bar\Pi_{s-}\times\\
	\tilde\Xi_{s-}^{-1}(dY_s-
	\tilde\Xi_{s}^{-1}(1-\tilde\Upsilon_{s}^*\tilde\Upsilon_{s})\,ds)
\end{multline*}
in the pure jump case.  Using the It\^o rules, one can verify that 
$\bar\Pi_t\bar\rho_t$ satisfies the same equation as $\tau_t$.  But 
$\tau_t$ is uniquely defined; hence we conclude that 
$\rho_t=(\bar\Pi_t/{\rm Tr}[\tau_t])\bar\rho_t$.  By taking the trace 
of the nonlinear filtering equations it is easily verified that 
${\rm Tr}[\rho_t]={\rm Tr}[\bar\rho_t]=1$ a.s.\ for all $t$, and hence 
a.s.\ $\rho_t=\bar\rho_t$.
\end{proof}

To conclude the section, we will now prove that the innovations problem
can be solved for the class of systems that we have considered under some
mild conditions on the controls.  It is likely that the innovations
problem can be solved under more general conditions; however, as we will
not need this result in the following, we restrict ourselves to the
following case for sake of demonstration.

\begin{proposition}[Innovations problem]
Let $\tilde\Upsilon_s$ and the matrix elements of $\tilde L_s$ and 
$\tilde H_s$ be c{\`a}dl{\`a}g semimartingales that are adapted to the 
filtration generated by the innovations process.  The the observations 
$Y_t$ and the innovations $\overline{Z}_t$ generate the same 
$\sigma$-algebras (and hence also the same von Neumann algebras.)
\end{proposition}

\begin{proof}
First, note that we can restrict ourselves to the case of diffusive 
observations.  In the case of counting observations the result is trivial:
as $Y_t-\overline{Z}_t$ is a continuous process, $Y_t$ can be completely 
recovered as the discontinuous part of $\overline{Z}_t$.  Moreover, 
$\overline{Z}_t$ is $Y_t$-measurable by construction.  Hence $Y_t$ and 
$\overline{Z}_t$ generate the same $\sigma$-algebras and there is no 
innovations problem.

Things are not so simple in the diffusive case.  Denote by $\Sigma_t^Y$
the $\sigma$-algebra generated by $Y_t$ up to time $t$, and similarly by
$\Sigma_t^{\overline{Z}}$ the $\sigma$-algebra generated by
$\overline{Z}_t$ up to time $t$.  The inclusion
$\Sigma_t^{\overline{Z}}\subset\Sigma_t^Y$ holds true by construction, so
we are burdened with proving the opposite inclusion
$\Sigma_t^Y\subset\Sigma_t^{\overline{Z}}$.  This is essentially an issue
of ``causal invertibility'': given only the stochastic process
$\overline{Z}_t$, can we find a map that recovers the process $Y_t$ in a
causal manner?  Clearly this would be the case \cite{LiS01} if the
nonlinear filtering equation has a unique strong solution with respect to
$\overline{Z}_t$; as we have already established that it has a unique
strong solution with respect to $Y_t$ the two solutions must then
coincide, after which we can recover $Y_t$ from the formula
$dY_t=d\overline{Z}_t+{\rm Tr}[(\tilde\Upsilon_t^*\tilde L_t+
\tilde\Upsilon_t\tilde L_t^*)\rho_t]\,dt$.  Our approach will be precisely 
to demonstrate the uniqueness of $\rho_t$ with respect to 
$\overline{Z}_t$.

To this end, consider the diffusive nonlinear filtering equation for 
$\rho_t$ given in Proposition \ref{pro:KSdiff}, where we now consider it 
to be driven directly by the martingale $\overline{Z}_t$ rather than by 
the observations $Y_t$.  Now introduce the following quantities: let $X_t$ 
be a vector which contains as entries all the matrix elements of $\rho_t$, 
$\tilde L_t$, $\tilde H_t$, and the process $\tilde\Upsilon_t$, and let 
$K_t$ be a vector that contains as entries all the matrix elements of 
$\tilde L_t$, $\tilde H_t$, the process $\tilde\Upsilon_t$, and 
$\overline{Z}_t$.  Then $K_t$ is a vector of semimartingales and we can 
rewrite the nonlinear filtering equation in the form
$$
	X_t=X_0+\int_0^tf(X_{s-})\,dK_s
$$
for a suitably chosen matrix function $f$.  By inspection, we see that
$f(X)$ is polynomial in the elements of $X$ and hence $f$ is a locally
Lipschitz function.  Thus there is a unique solution of the
nonlinear filtering equation with respect to $K_s$ up to an accessible
explosion time $\zeta$ \cite{Pro04}.  By uniqueness, the
nonlinear filtering solution $\rho_t$ with respect to $\overline{Z}_t$
must coincide with the solution with respect to $Y_t$ up to the explosion
time $\zeta$.  This also implies that for all $t<\zeta$, the solution with
respect to $\overline{Z}_t$ must be a positive unit trace matrix.  But the
set of all such matrices is compact, and hence the accessibility of
$\zeta$ is violated unless $\zeta=\infty$ a.s.  We conclude that the
unique solution $\rho_t$ with respect to $\overline{Z}_t$ exists for all
time.
\end{proof}

\section{A separation theorem}
\label{sec separation}

We are now finally ready to consider the control problem for quantum
diffusions; i.e., how do we choose the processes $L_t,H_t$, etc.\ in the
controlled quantum flow to achieve a certain control goal?  As we will be
comparing different control strategies, we begin by introducing some
notation which allow us to keep them apart.

\begin{definition}[Control strategy]
A control strategy $\mu=(\Xi,\Upsilon,S,L,H)$ is a collection of processes
$\Xi_t,\Upsilon_t,S_t,L_t,H_t$ defined on $[0,T]$ that satisfy the
conditions of Definition \ref{def:cdiff}.  Given $\mu$, we denote by
$\mu_t=(\Xi_t,\Upsilon_t,S_t,L_t,H_t)$ the controls at time $t$, by
$\mu_{t]}=(\Xi_{[0,t]},\Upsilon_{[0,t]},S_{[0,t]},L_{[0,t]},H_{[0,t]})$ 
the control strategy on the interval $[0,t]$, and similarly by $\mu_{[t}$ 
the control strategy on the interval $[t,T]$.
\end{definition}

Each control strategy $\mu$ defines a different controlled quantum flow.
To avoid confusion, we will label the various quantities that are derived 
from the controlled flow by the associated control strategy.  For example, 
$j_t^\mu(X)$ and $Y_t^\mu$ are the flow and observations process obtained 
under the control strategy $\mu$, etc.

Our next task is to specify the control goal.  To this end, we introduce a 
cost function which quantifies how successful a certain control strategy 
is deemed to be.  The best control strategy is the one that minimizes the 
cost.

\begin{definition}[Cost function]
Let $C^\mu$ be a process of positive operators, possibly dependent on 
the control strategy $\mu$, such that $C_t^\mu$ is affiliated to 
${\rm vN}(\mu_t,\mathcal{B})$, the algebra generated by the initial 
system and the control strategy at time $t$, for each $t\in[0,T]$.  Let 
$C_T\in\mathcal{B}$.  The total cost is defined by the functional
$$
	J[\mu] = \int_0^T j_t^\mu(C_t^\mu)\,dt + j_T^\mu(C_T). 
$$
$C_t^\mu$ and $C_T$ are called the running and terminal cost operators, 
respectively.
\end{definition}

Ultimately our goal will be to find, if possible, an optimal control 
$\mu^*$ that minimizes the expected total cost $\mathbb{P}(J[\mu])$.  Let 
us begin by converting the latter into a more useful form.
Using the tower property of conditional expectations, we have
$$
  \mathbb{P}(J[\mu]) = \mathbb{P}
  \left(\int_0^T 
	\mathbb{P}(j_t^\mu(C_t^\mu)|\mathcal{Y}^\mu_t)\,dt + 
  	\mathbb{P}(j^\mu_T(C_T)|\mathcal{Y}^\mu_t)
  \right)
$$
where $\mathcal{Y}_t^\mu$ is the algebra generated by $Y_s^\mu$ up to time 
$t$ (note that the conditional expectations are well defined as 
$j_t^\mu(C_t^\mu)$ is affiliated to $j_t^\mu(\mathcal{B})
\otimes\mathcal{Y}_t^\mu$.)  But then
$$
	\mathbb{P}(J[\mu]) = \mathbb{P}\left(
		\int_0^T \pi^\mu_t(C_t^\mu)\,dt + 
		\pi^\mu_T(C_T)
	\right),
$$
and we see that the expected cost can be calculated from the associated 
filtering equation only.  As the filter lives entirely in the commutative 
algebra $\mathcal{Y}_T^\mu$ we can now proceed, as in the previous 
section, by converting the problem into a classical stochastic problem.  
To this end, we use the spectral theorem to map the commutative quantum 
probability space $(\mathcal{Y}_T^\mu,\mathbb{P})$ to the classical space 
$L^\infty(\Omega^\mu,\Sigma^\mu_T,\nu^\mu,{\bf P}^\mu)$ 
(note that different control strategies may not give rise to commuting 
observations, and hence the classical space depends on $\mu$.)  Thus we 
obtain the classical expression
$$
	\mathbb{P}(J[\mu]) = \mathbb{E}_{\mathbf{P}^\mu}
	\left(\int_0^T \pi^\mu_t(C_t^\mu)\,dt + 
	\pi^\mu_T(C_T)\right).
$$
We will use the same notations as in the previous section for the 
classical stochastic processes associated to $L_t,H_t$, etc.  In addition, 
we denote by $\Sigma_t^\mu$ the $\sigma$-algebra generated by the 
observations $Y_s^\mu$ up to time $t$.

We have now defined the cost as a classical functional for an arbitrary 
control strategy.  In practice not every control policy is allowed, 
however.  First, note that in practical control scenarios only a limited 
number of controls are physically available; e.g.\ in the example of 
Hamiltonian feedback $H_t=u(t)H$ we can only modulate the strength $u(t)$ 
of a fixed Hamiltonian $H$, and we certainly cannot independently control 
every matrix element of $S_t,L_t$, etc.  Moreover we have expressed the 
(classical) cost in terms of the filter state $\pi_t^\mu(\cdot)$; hence we 
should impose sufficient regularity conditions on the controls so that we 
can unambiguously obtain the filtered estimate from the observations using 
the nonlinear filtering equation of the previous section.  To this end we 
introduce an admissible subspace of control strategies that are realizable 
in the control scenario of interest, and we require the solution of the 
optimal control problem to be an admissible control.

\begin{definition}[Admissible controls]
Define the admissible range $\mathscr{B}_t\subset\mathbb{R}\times
\mathbb{C}\times\mathcal{B}\times\mathcal{B}\times\mathcal{B}$ for every
$t\in[0,T]$.  A control strategy $\mu$ is admissible if 
$\mu_t\in\mathscr{B}_t$ a.s.\ (in the sense that 
$(\tilde\Xi_t,\tilde\Upsilon_t,\tilde S_t,\tilde L_t,\tilde H_t)\in
\mathscr{B}_t$ a.s.) for every $t\in[0,T]$ and $\mu$ has c{\`a}dl{\`a}g 
sample paths.  The set of all admissible control strategies is denoted by 
$\mathscr{C}$.
\end{definition}

\begin{remark}
Note that in order to satisfy Definition \ref{def:cdiff}, $\mathscr{B}_t$ 
should be chosen to be a bounded set such that the only admissible $S$ are 
unitary matrices and the only admissible $H$ are self-adjoint matrices.
\end{remark}

The optimal control problem is to find, if possible, an admissible 
control strategy $\mu^*$ that minimizes the expected total cost, i.e.\ to 
find a $\mu^*\in\mathscr{C}$ such that
$$
	\mathbb{P}(J[\mu^*]) = \min_{\mu\in\mathscr{C}}\mathbb{P}(J[\mu]).
$$
In principle $\mu^*_t$ could depend on the entire history of observations
up to time $t$.  This would be awkward, as it would mean that the
controller should have enough memory to record the entire observation
history and enough resources to calculate a (possibly extremely
complicated) functional thereof.  However, as the cost functional only
depends on the filter, one could hope that $\mu^*_t$ would only depend on
$\rho_t^\mu$, the solution of the nonlinear filtering equation at time
$t$.  This is a much more desirable situation as $\rho_t^\mu$ can be
calculated recursively: hence we would not need to remember the previous
observations, and the feedback at time $t$ could be calculated simply by
applying a measurable function to $\rho_t^\mu$.  A control strategy that
separates into a filtering problem and a control map is called a separated
strategy.

\begin{definition}[Separated controls]
An admissible control strategy $\mu\in\mathscr{C}$ is said to be separated
if there exists for every $t\in[0,T]$ a measurable map
$u_t^\mu:\mathcal{S}(\mathcal{B})\to\mathscr{B}_t$ such that 
$\mu_t=u_t^\mu(\rho_t^\mu)$, where $\rho_t^\mu$ is the matrix solution of 
the nonlinear filtering equation at time $t$ and 
$\mathcal{S}(\mathcal{B})$ is the set of positive matrices in 
$\mathcal{B}$ with unit trace.  The set of all separated admissible 
strategies is denoted by $\mathscr{C}^0$.
\end{definition}

The main technique for solving optimal control problems in discrete time
is dynamic programming, a recursive algorithm that operates backwards in
time to construct an optimal control strategy.  The infinitesimal form of
the dynamic programming recursion, Bellman's functional equation, provides
a candidate optimal control strategy in separated form.  The goal of this
section is to prove that if we can find a separated strategy
$\mu\in\mathscr{C}^0$ that satisfies Bellman's equation, then this
strategy is indeed optimal with respect to {\it all} control strategies in
$\mathscr{C}$, i.e.\ even those that are not separated. This result is
known as a separation theorem.  In classical stochastic control this 
result was established for linear systems in a classic paper by Wonham 
\cite{Won68} and for finite-state Markov processes by Segall \cite{Seg77}.  
The proof of the quantum separation theorem below proceeds along the same 
lines.

We begin by introducing the expected cost-to-go, i.e.\ the cost incurred 
on an interval $[t,T]$ conditioned on the observations up to time $t$.

\begin{definition}[Expected cost-to-go]\label{def:costtogo} 
Given an admissible control strategy $\mu$, the expected cost-to-go at 
time $t$ is defined as the random variable
$$
	W[\mu](t) = \mathbb{E}_{\mathbf{P}^\mu}
	\left(\left.
		\int_t^T  \pi^\mu_s(C_s^\mu)\,ds + 
		\pi^\mu_T(C_T) 
	\right|
		\Sigma^\mu_t
	\right).
$$
\end{definition}

The basic idea behind dynamic programming is as follows.  Regardless of 
what conditional state we have arrived at at time $t$, an optimal control 
strategy should be such that the expected cost incurred over the remainder 
of the control time is minimized; in essence, an optimal control should 
minimize the expected cost-to-go.  This is Bellman's principle of 
optimality.  To find a control that satisfies this requirement, one could 
proceed in discrete time by starting at the final time $T$, and then 
performing a recursion backwards in time such that at each time step the 
control is chosen to minimize the expected cost-to-go.  We will not detail 
this procedure here; see e.g.\ \cite{Kus71}.  
Taking the limit as the time step goes to zero gives the infinitesimal 
form of this recursion, i.e.\ Bellman's functional equation
$$
	-\frac{\partial V}{\partial t}(t,\theta) 
	=\min_{u\in \mathscr{B}_t}
	\left\{
		\mathscr{L}(u)V(t,\theta) + 
		{\rm Tr}[\theta\,\tilde C_t^{u}]
	\right\},\qquad
	t\in [0,T],~\theta\in\mathcal{S}(\mathcal{B})
$$
subject to the terminal conditon $V(T,\theta) = {\rm Tr}[\theta\,C_T]$ 
(recall that $C_t^\mu$ is affiliated to ${\rm vN}(\mu_t,\mathcal{B})$, and 
hence $\tilde C_t^\mu=\tilde C_t^{\mu_t}$ can be considered a 
$\mathcal{B}$-valued measurable function of $\mu_t$.)
The value function $V(t,\theta)$ essentially represents the expected 
cost-to-go conditioned on the event that the solution $\rho_t^\mu$ of the 
nonlinear filtering equation takes the value $\theta$ at time $t$.  If 
the minimum in Bellman's equation can be evaluated for all $t$ and 
$\theta$, then it defines a separated control strategy $\mu$ by
$$
	u_t^\mu(\theta) = 
	\mathop{\mathrm{argmin}}_{u\in\mathscr{B}_t}
	\left\{\mathscr{L}(u)V(t,\theta) + 
		{\rm Tr}[\theta\,\tilde C_t^{u}]
	\right\},
	\qquad
	t\in [0,T],~\theta\in\mathcal{S}(\mathcal{B}). 
$$
In these equations $\mathscr{L}(u)$ denotes the infinitesimal generator of 
the matrix nonlinear filtering equation given the control $u$; i.e., it 
is the operator that satisfies for any admissible control strategy $\mu$ 
the It\^o change of variables formula
$$
	f(t,\rho_t^\mu)=f(s,\rho_s^\mu)+\int_s^t\left\{
		\frac{\partial f}{\partial\sigma}(\sigma,\rho_\sigma^\mu)+
		[\mathscr{L}(\mu_\sigma)f]
			(\sigma,\rho_\sigma^\mu)
		\right\}\,d\sigma+
	\int_s^t G_{\sigma-}^{\mu,f}(\rho_{\sigma-}^\mu)\,d\overline{Z}_\sigma
$$
where $f$ is any sufficiently differentiable function ($C^2$ in the 
diffusive case, $C^1$ in the pure jump case.)  The expression for 
$\mathscr{L}(u)$ and $G_\sigma^{\mu,f}$ is standard \cite{Pro04} and can 
be obtained directly from Propositions \ref{pro:KSdiff} and 
\ref{pro:KSjump}.

Our brief discussion of dynamic programming is intended purely as a 
motivation for what follows.  Even if we had given a rigorous description 
of the procedure, the solution of Bellman's equation can only give a 
candidate control strategy and one must still show that this control 
strategy is indeed optimal.  Thus, rather than deriving Bellman's 
equation, we will now take it as our starting point and show that if we 
can find a separated control that solves it, then this control is optimal 
with respect to all admissible controls (i.e.\ there does not exist an 
admissible control strategy that achieves a lower expected total cost.)

\begin{theorem}[Separation theorem]\label{thm:optimality}
Suppose there exists a separated admissible control strategy $\mu \in
\mathscr{C}^0$ and a function $V:\ [0,T]\times \mathcal{S}(\mathcal{B})
\to \mathbb{R}$ such that
\begin{enumerate}
  \item\label{item smooth} The function $V$ is $C^1$ in the first variable 
	and $C^2$ in the second variable (diffusive case), or $C^1$ in 
	both variables (pure jump case).
  \item\label{eq Bellman V}
	For all $t\in[0,T]$ and $\theta\in\mathcal{S}(\mathcal{B})$, the 
	function $V$ satisfies 
        \begin{equation*}
        	\frac{\partial V}{\partial t}(t,\theta) + 
		\mathscr{L}(u_t^\mu(\theta))V(t,\theta) + 
		{\rm Tr}\left[\theta\,\tilde C_t^{u_t^\mu(\theta)}\right] 
		= 0. 
        \end{equation*}
  \item\label{eq optimality criterion}
	For all $t\in[0,T]$, $u\in\mathscr{B}_t$ and 
	$\theta\in\mathcal{S}(\mathcal{B})$, the function $V$ satisfies
        \begin{equation*}
        	\frac{\partial V}{\partial t}(t,\theta) + 
		\mathscr{L}(u)V(t,\theta) + 
		{\rm Tr}[\theta\,\tilde C_t^u]\ge 0.
	\end{equation*}
  \item\label{eq terminal condition}
	For all $\theta\in\mathcal{S}(\mathcal{B})$ the function $V$ 
	satisfies the terminal condition
        \begin{equation*}
        	V(T,\theta) = {\rm Tr}[\theta\,C_T].
	\end{equation*}	
\end{enumerate}
Then the separated strategy $\mu$ is optimal in $\mathscr{C}$, i.e.\
$\mathbb{P}(J[\mu])=\min_{\mu'\in\mathscr{C}}\mathbb{P}(J[\mu'])$.
\end{theorem}

\begin{proof}
We begin by showing that for the candidate optimal control $\mu$, the 
value function $V(t,\rho_t^\mu)$ evaluated at the solution of the 
nonlinear filtering equation at time $t$ equals the expected cost-to-go 
$W[\mu](t)$.  To this end, we substitute condition \eqref{eq Bellman V} 
with $\theta=\rho_t^\mu$ and the terminal condition \eqref{eq terminal 
condition} into Definition \ref{def:costtogo}.  This gives
$$
	W[\mu](t) = \mathbb{E}_{\mathbf{P}^\mu}
	\left(\left.
		V(T,\rho^\mu_T)-
		\int_t^T\left\{
			\frac{\partial V}{\partial s}(s,\rho^\mu_s)+ 
			[\mathscr{L}(u_s^\mu(\rho^\mu_s))V](s,\rho^\mu_s)
		\right\}ds 
	\,\right|
		\Sigma^\mu_t
	\right).
$$
The purpose of condition \eqref{item smooth} is to ensure that we can 
apply the It\^o change of variables formula to $V(t,\rho^\mu_t)$.  This 
gives
$$
	W[\mu](t) = \mathbb{E}_{\mathbf{P}^\mu}
	\left(\left.
		V(t,\rho^\mu_t)+\int_t^T G_{s-}^{\mu,V}(\rho_{s-}^\mu)\,
			d\overline{Z}_s
	\,\right|
		\Sigma^\mu_t
	\right).
$$
But by a fundamental property of stochastic integrals \cite{Pro04},
the stochastic integral of the bounded process $G_{t-}^{\mu,V}(\rho_{t-}^\mu)$
against the square-integrable martingale $\overline{Z}_t$ is itself 
a martingale.  Hence the conditional expectation of the second term 
vanishes, and as $\rho_t^\mu$ is $\Sigma_t^\mu$-measurable we obtain 
immediately
$$
	W[\mu](t) = V(t,\rho^\mu_t). 
$$
In particular, since $\rho^\mu_0=\rho$, we find that $V(0,\rho)$ equals 
the expected total cost
\begin{equation}\label{eq expected total cost}
	V(0,\rho) = \mathbb{P}(J[\mu]).
\end{equation}
To show that $\mu$ is optimal, let $\psi\in\mathscr{C}$ be an arbitrary
admissible control strategy.  We follow essentially the same argument as 
before, but now in the opposite direction.  Using the It\^o change of 
variable formula, we can verify that
$$
	V(t,\rho_t^\psi)=
	\mathbb{E}_{\mathbf{P}^\psi}
	\left(\left.
		V(T,\rho_T^\psi)-
		\int_t^T\left\{
			\frac{\partial V}{\partial s}(s,\rho^\psi_s)+ 
			[\mathscr{L}(\psi_s)V](s,\rho^\psi_s)
		\right\}ds 
	\,\right|
		\Sigma^\psi_t
	\right).
$$
Using condition \eqref{eq terminal condition} and the inequality \eqref{eq 
optimality criterion} with $\theta=\rho^\psi_t$ and $u=\psi_t$, we obtain
$$
	V(t,\rho_t^\psi)\le 
	\mathbb{E}_{\mathbf{P}^\psi}
	\left(\left.
		\pi_T^\psi(C_T)+
		\int_t^T \pi_s^\psi(C_s^{\psi})\,ds 
	\,\right|
		\Sigma^\psi_t
	\right)=W[\psi](t).
$$
But $\rho_0^\psi=\rho_0^\mu=\rho$, so together with Equation
\eqref{eq expected total cost} we obtain the inequality
$$
	\mathbb{P}(J[\mu]) = V(0,\rho) \le \mathbb{P}(J[\psi])
$$
for any $\psi\in\mathscr{C}$, which is the desired result.
\end{proof}

\begin{remark}
In a detailed study of the optimal control problem, separation theorems
are only a first step.  What we have not addressed here are conditions 
under which Bellman's equation can in fact be shown have a solution, nor 
have we discussed conditions on the feedback function $u_t^\mu$ of a 
separated control strategy that guarantee that the closed-loop system is 
well-defined (i.e., that it gives rise to c{\`a}dl{\`a}g controls.)
These issues have yet to be addressed in the quantum case.
\end{remark}

The simplicitly of the separation argument makes such an approach
particularly powerful.  The argument is ideally suited for quantum optimal
control, as we only need to compare the solution $V$ of a deterministic
PDE to the solution of the controlled quantum filter separately for every
control strategy.  Hence we do not need to worry about the fact that
different strategies give rise to different, mutually incompatible
observation algebras---the corresponding filters are never compared
directly.  The argument is readily extended, without significant
modifications, to a wide variety of optimal control problems: on a finite
and infinite time horizon, with a free terminal time or with time-average
costs (see e.g.\ \cite{OS05}).

Another option is to rewrite the cost directly in terms of the linear 
filtering equation, for example in the diffusive case:
\begin{multline*}
  \mathbb{P}(J[\mu]) = 
  \mathbb{P}
  \left(\int_0^T 
	U_t^{\mu*}C_t^\mu U_t^\mu\,dt + 
  	U_T^{\mu*}C_T U_T^\mu
  \right) \\
  =
  \mathbb{P}
  \left(\int_0^T 
	V_t^{\mu*}C_t^\mu V_t^\mu\,dt + 
  	V_T^{\mu*}C_T V_T^\mu
  \right)
  =
  \mathbb{R}^\mu\left(\int_0^T 
	\sigma_t^\mu(C_t^\mu)\,dt + 
  	\sigma_T^\mu(C_T)
  \right),
\end{multline*}
where $\mathbb{R}^\mu(X)=\mathbb{P}(U_T^\mu XU_T^{\mu*})$.  Note that 
under $\mathbb{R}^\mu$, the observation process $Y_t^\mu$ is a Wiener 
process.  Hence we can now perform dynamic programming, and find a 
separation theorem, directly in terms of the linear filtering equation.  
This is particularly useful, for example, in treating the risk-sensitive 
control problem \cite{Matt05}.

Finally, a class of interesting quantum control problems can be formulated 
using the theory of quantum stopping times \cite{PaSi87}; this gives rise 
to optimal stopping problems and impulse control problems in the quantum 
context.  Such control problems are explored in \cite{vH06} using a 
similar argument to the one used above.

\bibliographystyle{amsalpha}
\bibliography{ref}

\end{document}